\documentclass[aps,prd,showpacs,superscriptaddress,onecolumn,raggedfooter,raggedbottom]{revtex4-1}   

\usepackage{amssymb}
\usepackage{amsmath}
\usepackage{amssymb}
\usepackage{graphicx}
\usepackage{subfigure}
\usepackage{color}
\usepackage{mathrsfs} 
\usepackage{float}

\usepackage{soul}

\usepackage{hyperref}

\begin{document}

\title{Kink-Antikink Interaction Forces and Bound States in a $\phi^4$ Model with Quadratic and Quartic dispersion}

\author{G.~A.\ Tsolias}
\affiliation{Department of Mathematics and Statistics, University of Massachusetts,Amherst, MA 01003-4515, USA}

\author{Robert J.\ Decker}
\affiliation{Mathematics Department, University of Hartford, 200 Bloomfield Ave., West Hartford, CT 06117, USA}

\author{A.\ Demirkaya}
\affiliation{Mathematics Department, University of Hartford, 200 Bloomfield Ave., West Hartford, CT 06117, USA}

\author{T.J. Alexander}
\affiliation{Institute of Photonics and Optical Science (IPOS), School of Physics, The University of Sydney, NSW 2006, Australia}

\author{P.~G.\ Kevrekidis}
\affiliation{Department of Mathematics and Statistics, University of Massachusetts,Amherst, MA 01003-4515, USA}

\begin{abstract}
We consider the interaction of solitary waves in a model involving the well-known $\phi^4$ Klein-Gordon theory, but now bearing both Laplacian and biharmonic terms with different
prefactors. As a result of the competition of the respective linear operators, we obtain three distinct cases as we vary the model parameters.  
In the first the biharmonic effect dominates, yielding an oscillatory 
inter-wave interaction; in the third the harmonic effect prevails yielding exponential
interactions, while we find an intriguing linearly modulated exponential effect
in the critical second case, separating the above two regimes.
For each case, we calculate the force between the kink and antikink when initially separated with sufficient distance. Being able to write the acceleration as a function of the separation distance, and its corresponding ordinary differential equation, 
we test the corresponding predictions, finding very good agreement, where appropriate,
with the corresponding partial differential equation results. Where the two findings
differ, we explain the source of disparities. Finally, we offer a first glimpse of the interplay of harmonic and biharmonic effects on the results of kink-antikink
collisions and the corresponding single- and multi-bounce windows.\end{abstract}

\maketitle

\section{Introduction}

The study of nonlinear Klein-Gordon models is a topic that has a rich
history. Many of the early developments on the subject have focused on the
mathematically appealing theory of the inverse scattering transform and
integrable systems~\cite{ablowitz,ablowitz2,drazin}, such as the famous
sine-Gordon equation~\cite{gibbon,oursg}. However, more recently,
the intriguing features stemming from non-integrable dynamics have
been at the center of numerous studies centered around, e.g., the~$\phi^4$
model~\cite{ourp4}. The latter has often been considered to be a prototypical
 system for phase transitions, ferroelectrics, and high-energy physics
 among other themes~\cite{gibbon,ourp4}. Moreover, it has been 
 a central point of  both analytical and numerical explorations,
 involving kink interactions, collective coordinates, resonant dynamics (including
 with impurities)
 starting from the 1970's and extending
 over nearly 5 decades~\cite{kud,aub,get,AKL,Sugiyama,Campbell,belova,Ann,goodman,goodman2,weigel,weigel2} and even reaching to
 this day~\cite{weig1,clisthenis,bazeia}; see also the recent
 recap of~\cite{lizunova}. 

On the other hand, more recently, a diverse set of variants of the so-called
nonlinear beam (or biharmonic) wave equation have been considered also;
a collection of relevant examples includes, 
e.g.,~\cite{levandosky,champneys,CM,karageorgis}. 
The corresponding models also span a diverse array of contexts, including, e.g.,
suspension bridges and the propagation of traveling waves therein.
Another setting that has emerged very recently and has substantially promoted
the relevance of biharmonic models has been the form of generalized nonlinear
Schr{\"o}dinger (NLS) type models in nonlinear optics in the context of the so-called
``pure-quartic solitons''~\cite{pqs}. Not only has this type of dispersion
engineering been realized in the lab, but it has also been used in the context
of the so-called pure-quartic soliton laser~\cite{pqs3}. 
Mathematical studies of the existence and stability of solitary waves of such
equations are also ongoing~\cite{beam_demirkaya,beam1,atanas}. 

It has, however, been
recognized that the typical scenario involves (e.g., in the NLS experimental
settings discussed above) both regular quadratic and quartic dispersion 
and that experimental settings have the ability to tune the interplay between
the two~\cite{pqs2}. This results in the form of a {\it generalized} NLS equation
incorporating both quadratic and quartic dispersion and thus presenting the potential
for engineering a situation involving competition between the two~\cite{pqs2}. While the generalized
NLS setting is the most canonical one to consider in the realm of nonlinear
optics, in the present work we opt to consider the slightly simpler, yet highly
informative, setting of a corresponding Klein-Gordon model. The rationale
behind the latter choice involves the fact that the two models {\it share} the
same existence properties, at least in one spatial dimension, yet the nature
of the real field-theory renders the analytical calculations somewhat simpler,
especially as regards the stability and dynamical implications of the inter-wave
interactions. 

Our aim in the present work is to formulate the existence problem of a single
kink in a model incorporating quadratic and quartic dispersion in the presence
of a $\phi^4$ potential (this part will be entirely analogous to the corresponding
generalized NLS case). Subsequently, we intend to explore the interaction of two such waves
and identify their pairwise interaction force and how it depends on the model
parameters. Subsequently, conclusions of the analytical theory will be tested
against full numerical computations of the interaction dynamics. Lastly, collisions between two coherent structures will be simulated, and the possible scenarios thereof will be considered. Our aim is to reveal the possibility that either
the biharmonic effect may dominate (yielding oscillatory tails and forces,
equilibrium steady states of alternating stability etc.) or the harmonic effect
will prevail (featuring exponential interactions and forces). The critical case
between the two and its own intriguing behavior will be revealed as well. 
In our study of collisions and, in particular, in the case kinks and antikinks
interact and eventually separate,
we create velocity-out versus velocity-in curves. These curves show windows of velocity-in values  for which we see different numbers of bounces before the 
coherent structures separate. We compare this behavior to both the ``pure $\phi^4$'' case of Eq. \ref{phi4} and the ``pure biharmonic'' special-case limits of the 
present model interpolating between them.

\section{Model Setup \& Kink-Antikink Tail Behavior}

The standard $\phi^4$ Klein-Gordon theory yields the field equation
\begin{equation}
u_{tt}=u_{xx}-V'(u)
\label{phi4}
\end{equation}
where $V(u)=\frac{1}{2}(u^{2}-1)^{2}$. In \cite{beam1,manton-decker}, a variant of this equation was explored where the harmonic spatial derivative term was replaced by a biharmonic term of the form:
\begin{equation}
u_{tt}=-u_{xxxx}-V'(u).
\label{beam}
\end{equation}
Here, as indicated in the above section, motivated by the corresponding
generalized NLS of~\cite{pqs2}, we explore a model incorporating
the competition of the features of the two models:
\begin{equation}
u_{tt}=\alpha u_{xx}-\beta u_{xxxx}-V^{\prime }(u)
\label{par-phi4}
\end{equation}
where $\alpha $ and $\beta $ are assumed
positive (to ensure the competition referred to above) and the potential function $V(u)$ is taken as before.  
When we pick $\alpha =1$ and $\beta =0$, we get Eq.~(\ref{phi4}) and when we pick $\alpha =0$ and $\beta =1$, we get Eq.~(\ref{beam}). Notice that while one of the coefficients could be scaled out
via a rescaling of space, we maintain both coefficients, in order to maintain the
tractability of the special case limits of $(0,1)$ and $(1,0)$, i.e., biharmonic
and harmonic respectively.

A central consideration of the present work is to explore both the features
of a single solitary wave, but also to examine the interaction between two
such waves, a kink and an antikink. 
We will use a method developed by Manton (as in \cite{Manton_nuclear,manton-decker}) to find the force between a separated kink and antikink as a function of the separation distance. To do this we must first determine the tail behavior for a single kink or antikink. Once the force is determined, we can use the corresponding acceleration to generate an ordinary
differential equation (ODE), whose behavior can then be compared to the soliton trajectories of Eq. (\ref{par-phi4}), i.e., the corresponding partial differential equation (PDE). 
As long as the separation distance between kink and antikink remains sufficiently large, the agreement between ODE and PDE should be quite good. However, in cases where the kink and antikink approach each other at distances comparable to their respective widths,
then it is no longer obvious that the ODE model should be an adequate description of
the full PDE dynamics and the exchanges of energy between the different modes
present in the latter~\cite{ourp4}. We will explore both the former
agreement (at large distances) and the latter deviations (at short ones)
in the numerical results below.
In order to determine the tail behavior of a single kink we proceed as follows. Substituting $\phi(x)=u(t,x)$ into Eq. (\ref{par-phi4}) we get the steady-state equation
\begin{equation}
    \alpha \phi''-\beta \phi^{(iv)}-V'(\phi)=0
    \label{steadyStateKink}
\end{equation}
To examine the relevant asymptotics, we substitute $\phi =1-\varepsilon e^{\lambda x}$ 
into the Eq. (\ref{steadyStateKink}). Neglecting
terms of $\varepsilon ^{2}$ and higher (i.e., linearizing),
for the above mentioned $\phi^4$ potential, we get%
\begin{equation}
-\alpha \lambda ^{2}+\beta \lambda ^{4}+4=0 
\label{lambdaEquation}
\end{equation}%
It is easy to show that the roots of this equation are real for $\alpha
\geq 4\sqrt{\beta }$ and complex for $\alpha <4\sqrt{\beta }$. In particular, for \(\alpha<4\sqrt{\beta}\):
\[\lambda_1 = \frac 1 2 \sqrt{\frac{4\sqrt \beta + \alpha} \beta } + i \frac 1 2 \sqrt{\frac{4\sqrt \beta - \alpha} \beta },\;\;\; \lambda_2 = \frac 1 2 \sqrt{\frac{4\sqrt \beta + \alpha} \beta } - i \frac 1 2 \sqrt{\frac{4\sqrt \beta - \alpha} \beta }\]
\[\lambda_3 = -\frac 1 2 \sqrt{\frac{4\sqrt \beta + \alpha} \beta } + i \frac 1 2 \sqrt{\frac{4\sqrt \beta - \alpha} \beta },\;\;\;\lambda_4 = -\frac 1 2 \sqrt{\frac{4\sqrt \beta + \alpha} \beta } - i \frac 1 2 \sqrt{\frac{4\sqrt \beta - \alpha} \beta }\]
for \(\alpha = 4 \sqrt{\beta}\) (critical case), the degenerate roots are:
\[\lambda_{1,2} =  \sqrt{\frac \alpha {2 \beta} },\;\;\; \lambda_{3,4} = - \sqrt{\frac \alpha {2 \beta} }\]
and for \(\alpha > 4 \sqrt{\beta}\):
\[ \lambda_1 =\sqrt{ \frac{\alpha-\sqrt{\alpha^2 -16 \beta}} {2\beta} }, \;\;\; \lambda_2 =\sqrt{ \frac{\alpha+\sqrt{\alpha^2 -16 \beta}} {2\beta} } \]
\[ \lambda_3 =-\sqrt{ \frac{\alpha-\sqrt{\alpha^2 -16 \beta}} {2\beta} }, \;\;\; \lambda_3 =-\sqrt{ \frac{\alpha+\sqrt{\alpha^2 -16 \beta}} {2\beta} } \]

For real $%
\lambda $, similarly to the pure $\phi^4$ case, our model for the tail behavior is \begin{equation} be^{-ax}+de^{-cx},
\label{realTail}\end{equation}
for the critical case with the double roots the model is 
\begin{equation} be^{-ax}(x-d) \label{criticalTail}\end{equation} (accounting for the relevant generalized eigenvector)
and for the complex $%
\lambda $ case (similarly to the pure biharmonic one), 
the model is 
\begin{equation}
be^{-ax}\cos (c(x-d)).
\label{complexTail}
\end{equation} 
We also know that in the real case $a=\lambda_1$ 
\(= \sqrt{ \frac{\alpha-\sqrt{\alpha^2 -16 \beta}} {2\beta} } \) and $c=\lambda_2 
$  \(= \sqrt{ \frac{\alpha+\sqrt{\alpha^2 -16 \beta}} {2\beta} } \) 
and in the complex case $a={\rm Re}(\lambda )$ 
\(  = \frac 1 2 \sqrt{\frac{4\sqrt \beta + \alpha} \beta }\) 
and $c={\rm Im}(\lambda )$ 
\(  = \frac 1 2 \sqrt{\frac{4\sqrt \beta - \alpha} \beta }\), while in the critical case \(a=\sqrt{\frac \alpha {2 \beta} }\). We use
curve fitting to get the other parameters 
\(b\) and \(d\) \color{black}. The results are in Table \ref{tail} for
a sequence of prototypical case examples that we have considered.

\begin{table}[H]
\centering
\begin{tabular}{c c c c}
$\alpha$ & $\beta$ & $ \lambda$ & Tail behavior \\ 
0 & 1 & 1-1i & $0.9700e^{-x}\cos (x-0.4083)$ \\ 
1 & 1 & 1.1180-0.8660i & $1.205e^{-1.118x}\cos (0.8660(x-0.9909))$ \\ 
2 & 1 & 1.2247-0.7071i & $1.793e^{-1.225x}\cos (0.7071(x-1.824))$ \\ 
3 & 1 & 1.3229-0.5000i & $3.614e^{-1.323x}\cos (0.5(x-3.299))$ \\ 
3.5 & 1 & 1.3693-0.3536i & $6.662e^{-1.369x}\cos (0.3536(x-4.942))$ \\ 
4 & 1 & 1.4142 & $3.363e^{-1.414x}(x - 0.9786)$ \\ 
4.5 & 1 & 1.1042, 1.8113 & $4.451e^{-1.104x}-12.92e^{-1.811x}$ \\ 
5 & 1 & 1, 2 & $3.354e^{-x} - 25.64 e^{-2x}$ \\ 
6 & 1 & 0.8740, 2.2882 & $2.679e^{-0.8740x}-157.4e^{-2.288x}$ \\ 
1 & 0 & 2 & $2e^{-2x}$%
\end{tabular}
\caption{Single Kink Tail Behavior for different model parameter
$(\alpha,\beta)$ in columns 1 and 2. Column 3 yields the corresponding
(spatial) eigenvalues and column 4 the functional form providing the optimal fit to the tail behavior. {One can read off the values of $a$, $b$, $c$, $d$ in column 4 by referring to Equations (\ref{realTail}), (\ref{criticalTail}), (\ref{complexTail}).}
\label{tail}}.
\end{table}

\begin{figure}[h!]
     \includegraphics[width=0.29\textwidth]{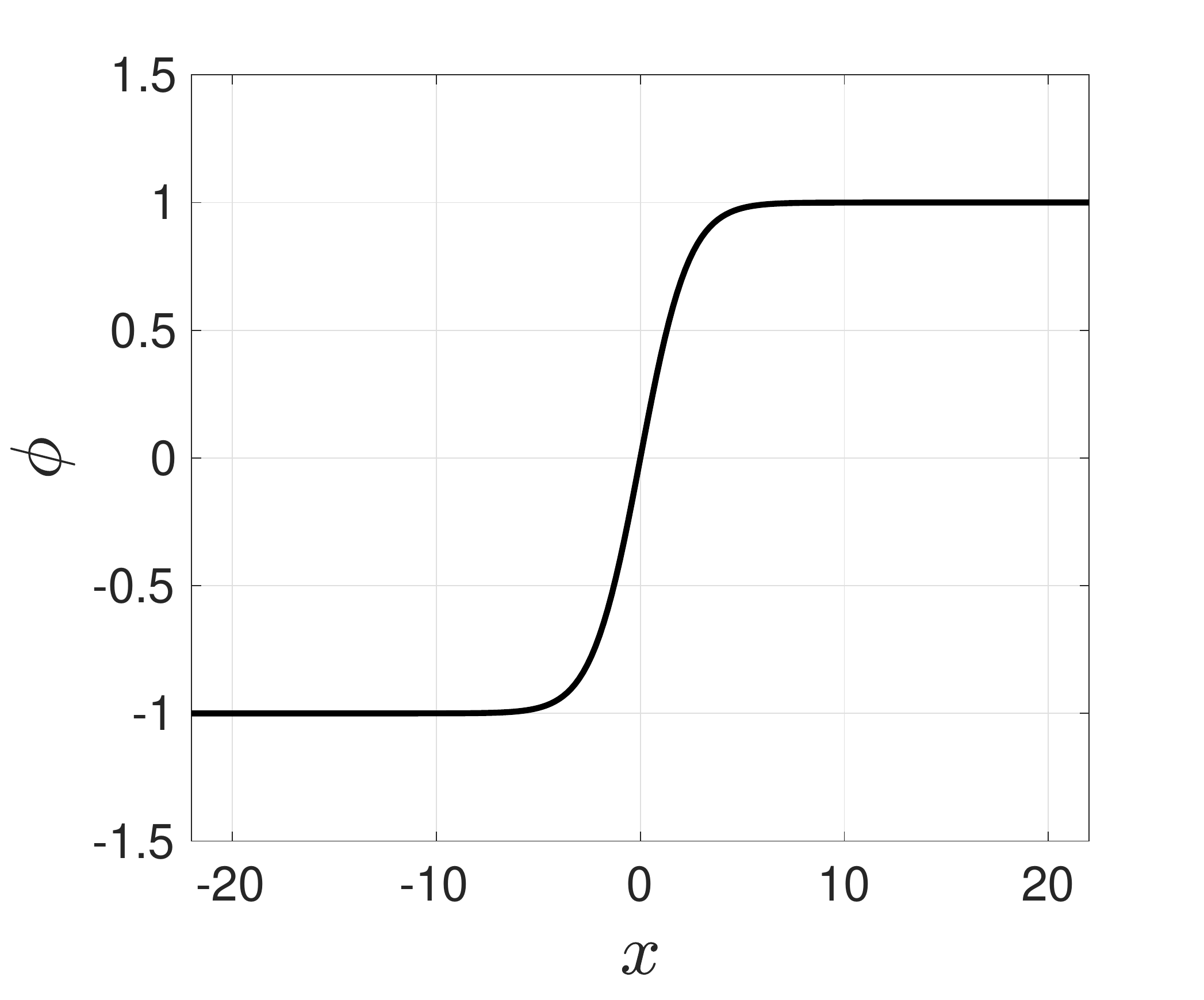}
         \includegraphics[width=0.29\textwidth]{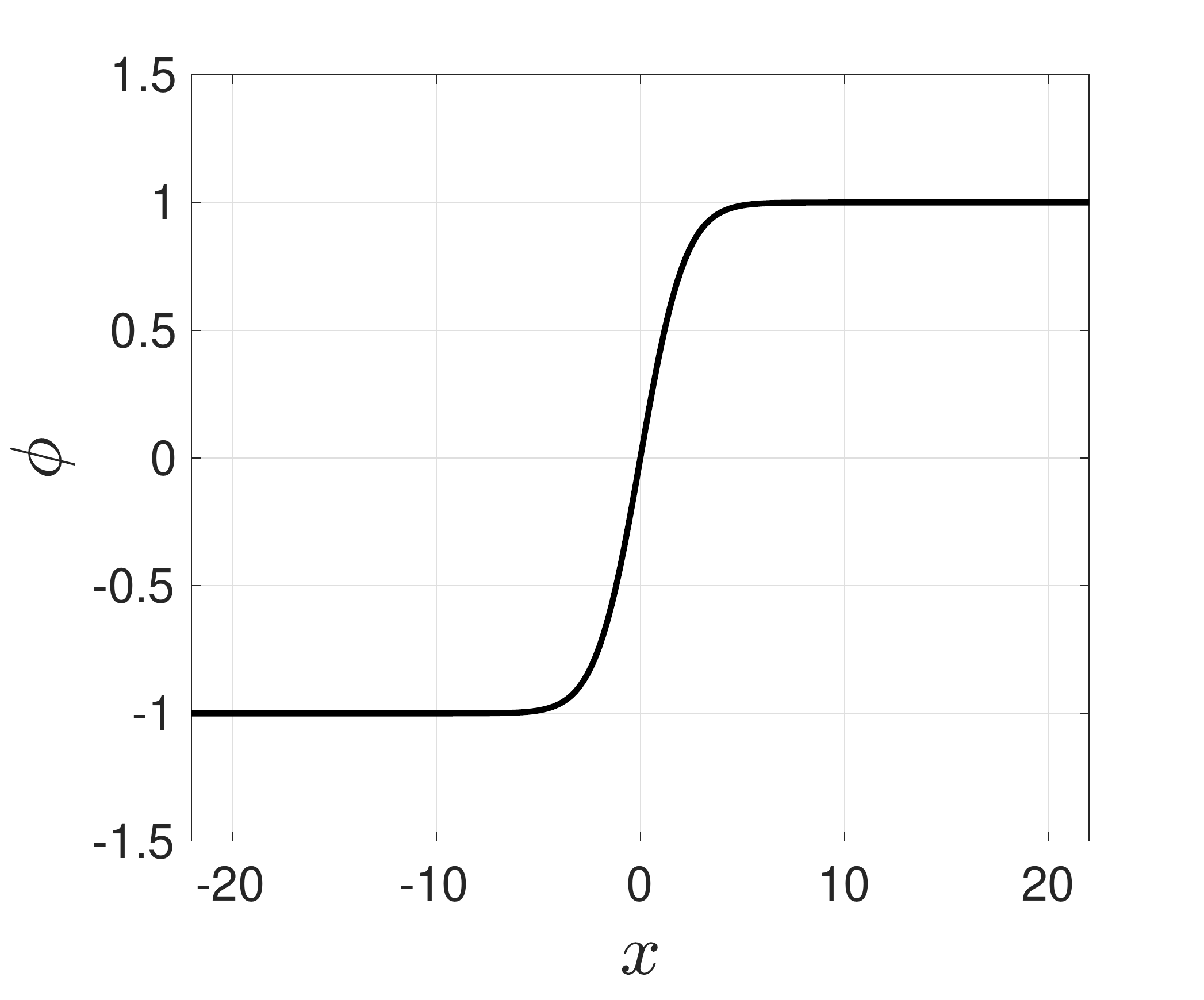}
         \includegraphics[width=0.29\textwidth]{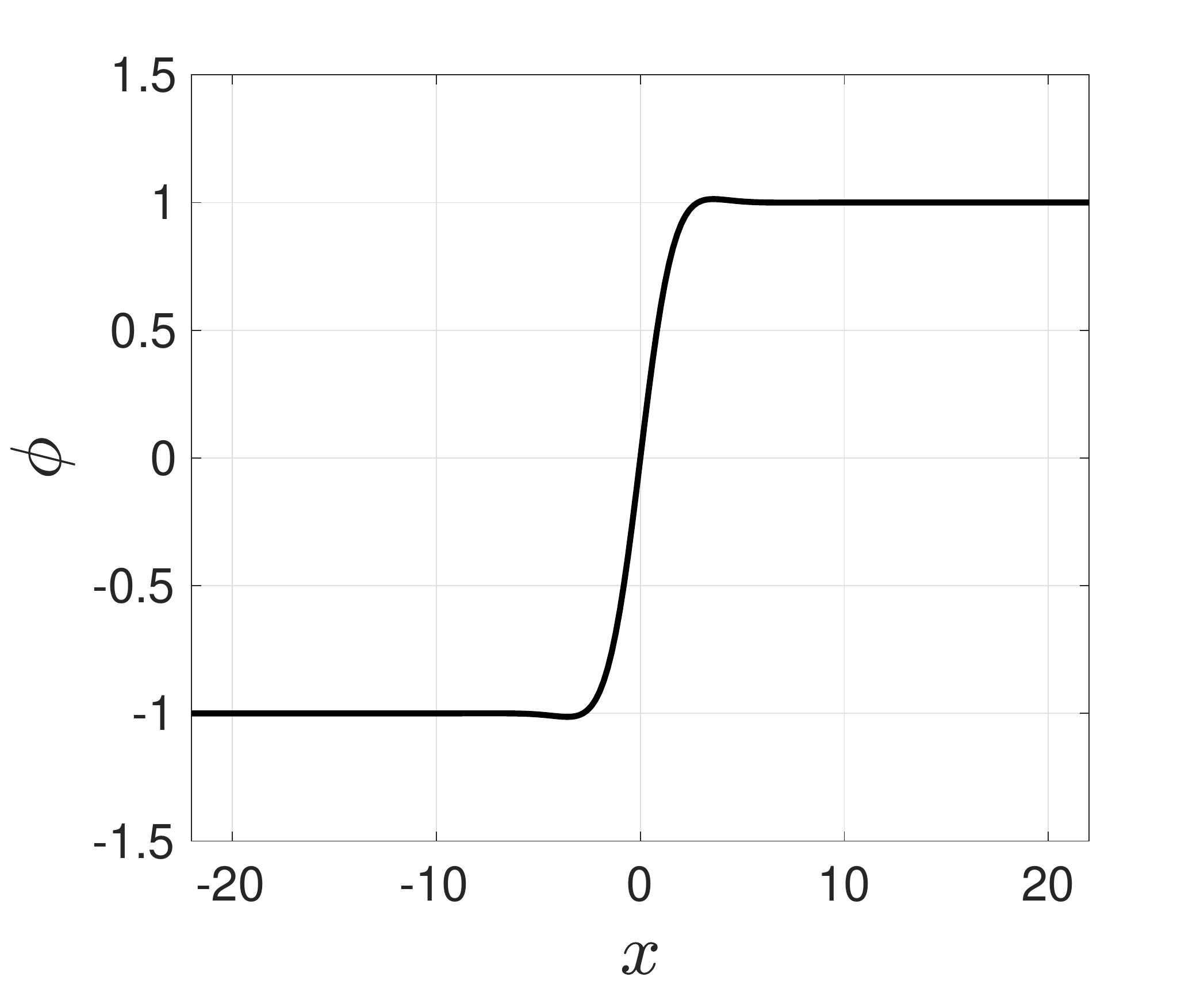} 
         
         \includegraphics[width=0.29\textwidth]{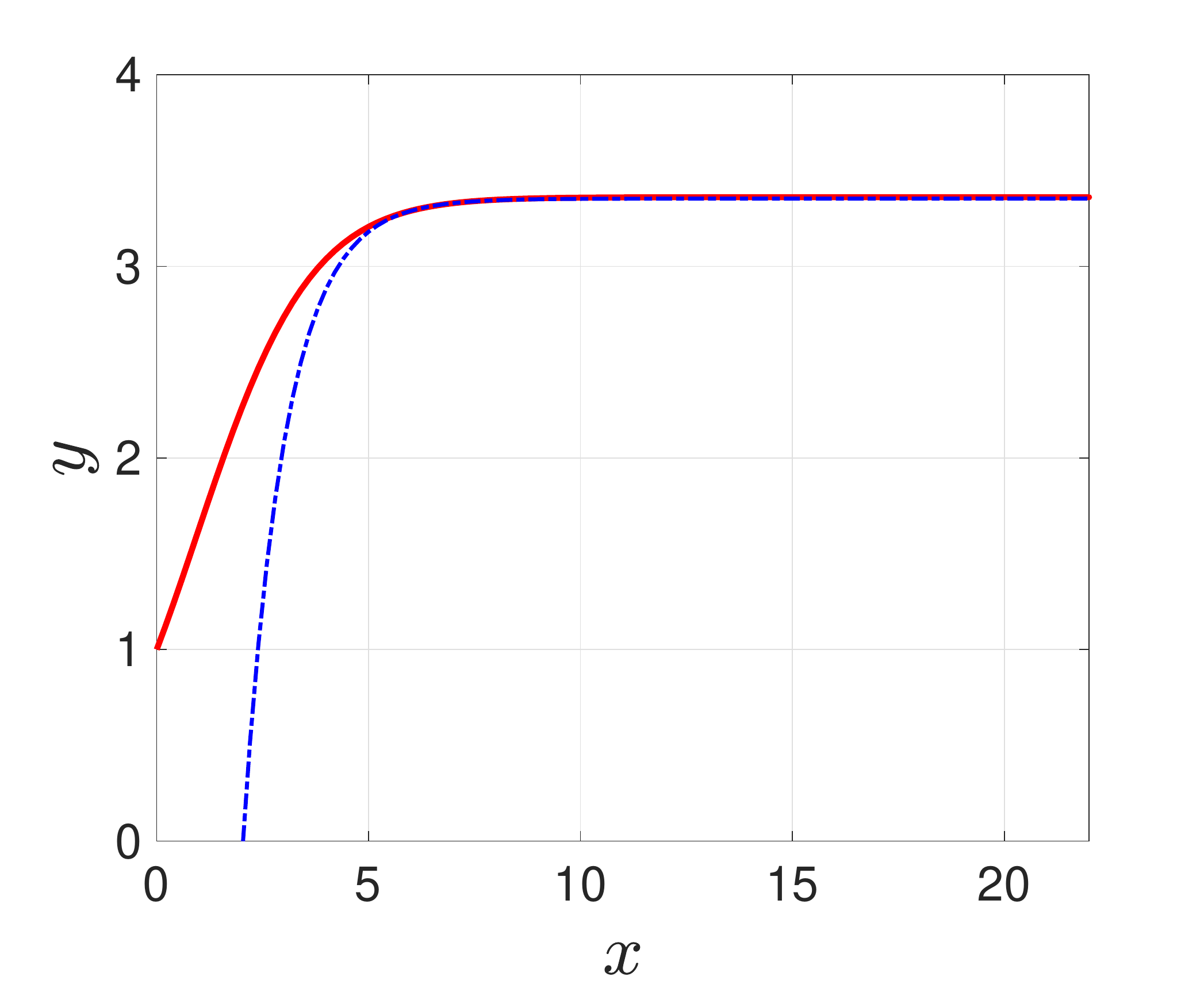}
         \includegraphics[width=0.29\textwidth]{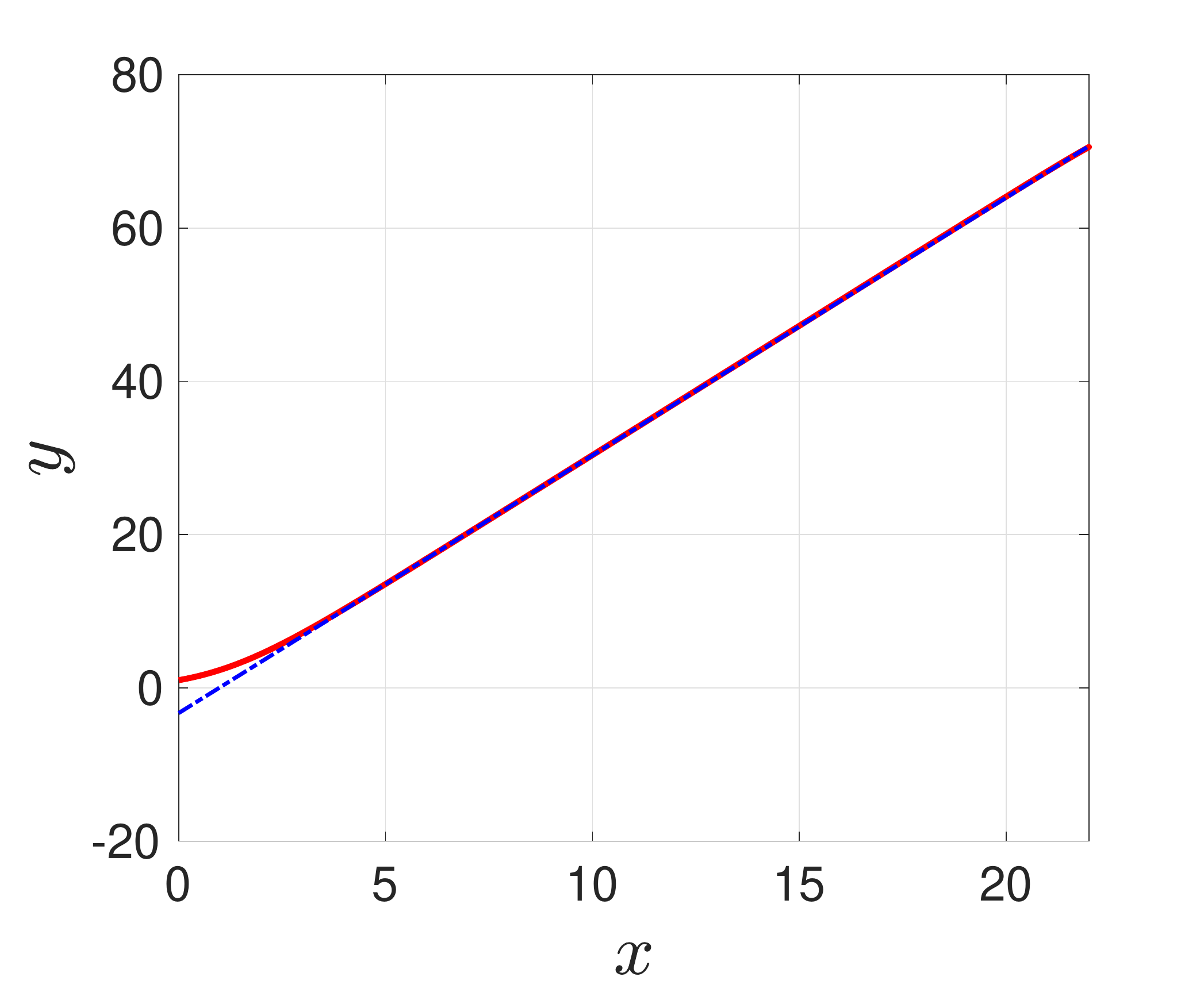}
         \includegraphics[width=0.29\textwidth]{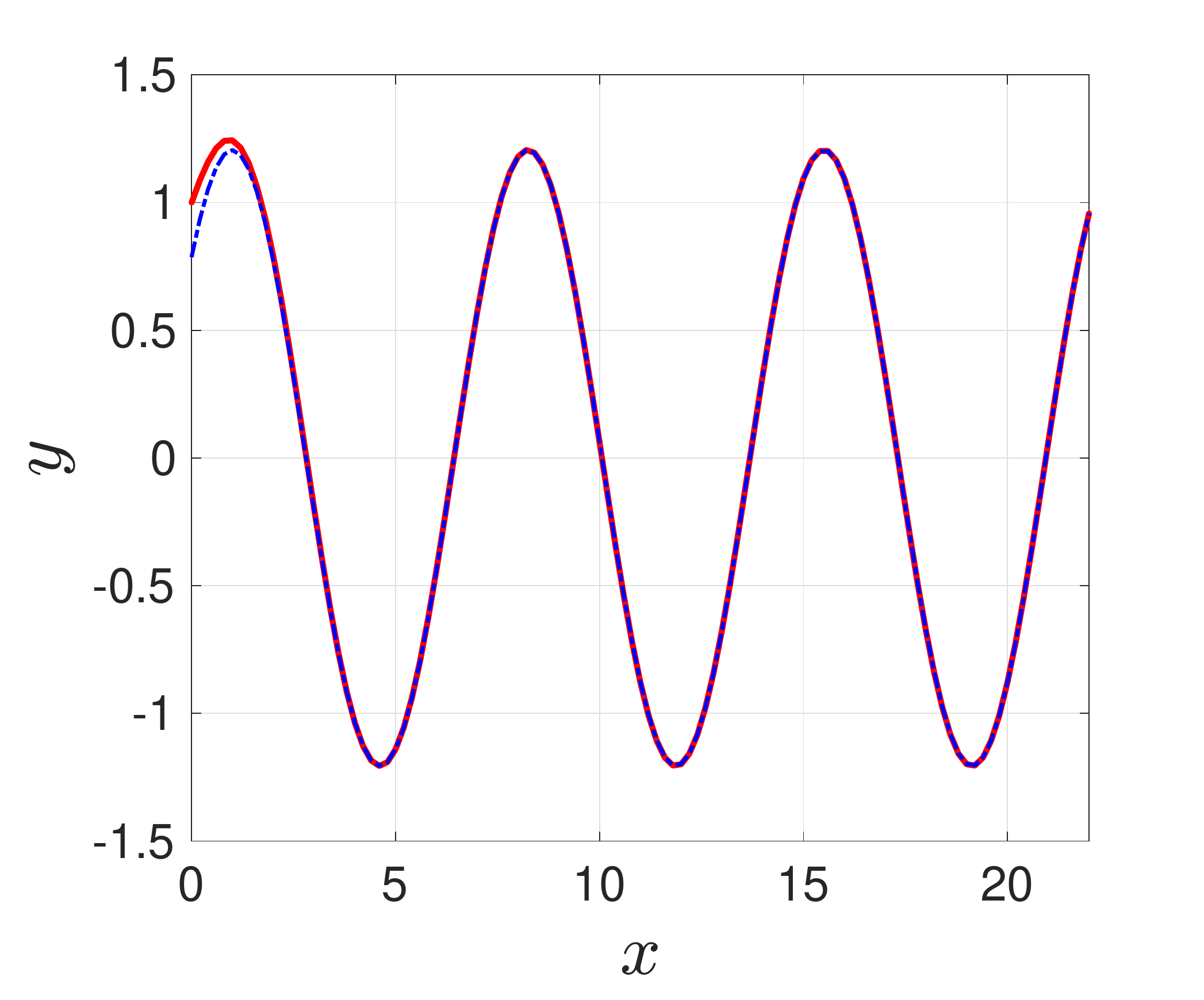}
\caption{The top three panels show a single kink, $\phi_K(x)$, for each of the cases $\alpha=1$, $\alpha=4$, $\alpha=5$ and $\beta=1$ (respectively, left to right). The bottom three panels illustrate the behavior of the right-side tails of these kinks by graphing $1-\phi_K(x)$ multiplied be $\exp(k x)$, where $k$ (positive) is Re$(\lambda)$ in the complex case, $\lambda$ in the critical case, and the smaller (in absolute value) $\lambda$ in the real case.  Superimposed are the fitted curves, also multiplied by $\exp(kx)$. In all cases the red solid curve is $\exp(k x)(1-\phi_K)$. The blue dash-dot curve is the fitted equation multiplied by $\exp(kx)$; the specific equations for each case are $y=1.205\cos(0.8660(x-0.9890))$, $y=3.274(x-0.8606)$ and $y=3.308- 14.9e^{-x}$ (right to left respectively)}.
\label{kink_tail}
\end{figure}

In Figure \ref{kink_tail} we show a single kink, $\phi_K(x)$, in the top three panels and the tail-behavior of each kink in the bottom three panels for the cases $\alpha=1$, $\alpha=4$, $\alpha=5$ with $\beta=1$ (respectively, left to right). For the tail behavior we graph the right tail of $1-\phi_K(x)$ multiplied by $\exp(k x)$ as well as a model fitted to the tail of $1-\phi_K(x)$, also multiplied by $\exp(k x)$ (appropriate for $x$ sufficiently large). The value of $k$ is equal to the real part Re$(\lambda)$ in the complex case, $\lambda$ in the critical case, and the smaller (in absolute value) $\lambda$ in the real case (corresponding to the slow decay). 
We can observe an excellent agreement in the oscillatory case (especially factoring
in that we have multiplied the expression by an exponential, hence any deviation 
in the exponent would lead to an exponential growth). Similarly, also a remarkable
fit can be discerned even in the critical case, revealing the underlying linear dependence modulating 
the exponential decay of the generalized
eigenvector in this setting. The exponential case (originally doubly
exponential turned into a single exponential upon multiplication
by $\exp(k x)$) is found to be less accurate. {In the latter case of two real $\lambda$ an improved fit can be obtained to the model $b e^{-a x}+de^{-c x}$ if $c$ is left as a free parameter in the curve-fitting process (rather than using the value specified above which results from Eq. (\ref{lambdaEquation})).} It is an interesting
question for future work, whether a weakly nonlinear theory can
capture more accurately the correction to the leading exponential
dependence; however, for our present purposes, the current prediction
capturing adequately the leading order exponential tail 
behavior will suffice.

\section{THE FORCE BETWEEN KINK AND ANTIKINK}

In order to find the force or acceleration between kink and
antikink, we use the approach of Manton as in \cite{Manton_nuclear,manton-decker}; see also \cite{manton-decker} for details of the calculation in the case where $\alpha=0$ and $\beta=1$. We now briefly review some of the details of this force calculation. Consider the momentum $P=-\int_{x_1}^{x_2} u_t u_x dx$ on the interval $[x_1,x_2]$ ($P$ is conserved when the integral is over the entire real line). Differentiating under the integral and using Eq. (\ref{par-phi4}) we find that the force is given by 
\begin{equation}
\frac{dP}{dt}=F=\left[-\frac{1}{2}u_t^2 -\alpha \frac{1}{2}u
_{x}^{2}+\beta u _{x}u _{xxx}-\beta \frac{1}{2}u _{xx}^{2}+V(u )%
\right] _{x_{1}}^{x_{2}}.
\label{eqForce}
\end{equation}

For a field configuration
that is static or almost so, we can ignore the 
first term in the right hand side bracket of 
Eq.~(\ref{eqForce}). We consider a configuration 
\begin{equation}
    u(t,x)=\phi(x)=\phi_K(x+X(t))+\phi_{AK}(x-X(t))-1,
    \label{kink_antikink_combo}
\end{equation}
where $-X(t)$ is the position of the kink and $X(t)$ is the position of the antikink, which represents a kink-antikink pair approaching each other (as $t$ gets larger). Then, set $\eta=1-\phi$, $\eta_K=1-\phi_K$ and $\eta_{AK}=1-\phi_{AK}$ (where $\phi_{AK}$ is an antikink solution to Eq. (\ref{steadyStateKink})). Evaluating Eq. (\ref{eqForce}) from $x_1=x$ to $x_2 \rightarrow \infty$ results in
\begin{equation} 
F=\alpha (\eta _{K})_{x}(\eta
_{AK})_{x}-\beta (\eta _{K})_{x}(\eta _{AK})_{xxx}-\beta (\eta
_{K})_{xxx}(\eta _{AK})_{x}+\beta (\eta _{K})_{xx}(\eta _{AK})_{xx}-4(\eta
_{K})(\eta _{AK}),
\label{MantonForce}
\end{equation}
where we assume that the kink-antikink separation $2X$ is large. We have also assumed that $|x|<<X$ and have used the approximation $V(\phi)=V(1-\eta) \approx 2\eta^2$. Note that only the cross terms are left at this point. 

For the \textbf{complex case} the model for the tail behavior is 
$\eta_K=b e^{-ax} \cos(c(x-d)$, appropriate for $x$ sufficiently large. 

{Carrying out the derivatives in Eq. (\ref{MantonForce}) and  using  \(a={\rm Re}(\lambda ) = \frac 1 2 \sqrt{\frac{4\sqrt \beta + \alpha} \beta }\) 
and \(c={\rm Im}(\lambda )  = \frac 1 2 \sqrt{\frac{4\sqrt \beta - \alpha} \beta }\)
we get the following expression for the force:}
\begin{equation}
F =  - 2 \sqrt {\frac {16\beta -\alpha^2}{\beta}}  b^2 e^{-X \sqrt{\frac{4\sqrt \beta + \alpha} \beta }}   \cos \left( \sqrt{\frac{4\sqrt \beta - \alpha} \beta } \left(X-d\right) +\theta \right). \end{equation}
Here, we have that  \(\theta\in[0,\frac{\pi}{2}]\) such that  \( \tan \theta = \frac{\alpha}{\sqrt{16\beta- \alpha^2}}\). 

For the \textbf{critical case} the model for the tail behavior is \(b e^{-ax} (x-d) \). This, upon substituting \( a = \sqrt{\frac{\alpha}{2\beta}}\) and \(\alpha = 4 \sqrt \beta\) results in the force formula:
\begin{equation}
 F =-8 b^2 \sqrt{2 \alpha} e^{-2 a X} \left(X-\frac{\sqrt{2 \alpha}}{4}-d\right),
 \label{forceCritical}
\end{equation}
once again featuring a functional form reminiscent of that of
the kink tail.

For the \textbf{real case} the tail behavior is $b e^{-ax}+de^{-cx}$ 
and the force becomes
\[F=-b^{2}e^{-2Xa}\left( a^{2}\alpha -3a^{4}\beta +4\right)  -d^{2}e^{-2Xc}\left( c^{2}\alpha -3c^{4}\beta +4\right),
\bigskip \]
where: 
\begin{equation}
a = \sqrt{ \frac{\alpha-\sqrt{\alpha^2 -16 \beta}} {2\beta} } \;\;\; \textnormal{and} \;\;\; c= \sqrt{ \frac{\alpha+\sqrt{\alpha^2 -16 \beta}} {2\beta} }. \label{a_and_b}\end{equation}
Using these values the force formula can be written as
\begin{equation}F= \frac{\alpha ^2-16 \beta -\alpha  \sqrt{\alpha ^2-16 \beta }}{\beta } b^{2}e^{-2Xa}  + \frac{\alpha ^2-16 \beta+\alpha  \sqrt{\alpha ^2-16 \beta } }{\beta } d^{2}e^{-2Xc}.
\end{equation}
Notice that the coefficient of the slow term is always negative while the coefficient of the fast term is always positive.
\color{black}
Dividing the above formulae for the force by the mass gives the results in
the acceleration column of Table \ref{accelTable}. The values for $b$
and $d$ are determined by curve fitting the tail of a single kink, and are
shown in Table \ref{tail}. 

\begin{table}
\begin{tabular}{c c c c}
$\alpha$ & $\beta$ & Mass & Acceleration  \\ 
0 & 1 & 1.1852 & $6.351e^{-2x}\cos (2x-0.8166)$\\ 
1 & 1 & 0.9540 & $11.79e^{-2.236x}\cos (1.732x-1.464)$  \\ 
2 & 1 & 0.8052 & $27.66e^{-2.4494x}\cos (1.4142x-2.0559)$ \\ 
3 & 1 & 0.7031 & $98.30e^{-2.6458x}\cos (x-2.451)$ \\ 
3.5 & 1 & 0.6633 & $259.1e^{-2.7386x}\cos (0.7072x-2.429)$ \\ 
4 & 1 & 0.6290 & $407.0 e^{-2.828 x} (x -1.686)$  \\ 
4.5 & 1 & 0.5991 & $166.2 e^{-2.208 x} -3769 e^{-3.623 x}$ \\ 
5 & 1 & 0.5728 & $117.8 e^{-2 x} -27545 e^{-4 x}$ \\ 
6 & 1 & 0.5287 & $92.75e^{-1.748x}-2194577e^{-4.576x}$ \\ 
1 & 0 & 4/3 & $24e^{-4x}$ 
\end{tabular}
\caption{Mass and acceleration as a function of the half-separation distance $x$ of the kink and antikink. \label{accelTable}}
\end{table}

Next, we integrate the expressions for the force on a kink to get the potential energy for each case, and also find all fixed points and their stability type. 
For the potential function in the complex case we have
\begin{equation}U =  -  b^2   \sqrt {\frac {16\beta -\alpha^2}{2 \sqrt \beta}} e^{-X \sqrt{\frac{4\sqrt \beta + \alpha} \beta }}   \cos \left( \sqrt{\frac{4\sqrt \beta - \alpha} \beta } \left(X-d\right) +\vartheta \right)
\label{complexPotential}  \end{equation}
for \(\vartheta\in[0,\frac{\pi}{2}]\) such that  \( \tan \vartheta = \frac{\sqrt{4\sqrt\beta+ \alpha}}{\sqrt{4\sqrt\beta- \alpha}}\). 
For saddle points we have
\[
X^*(k) = d + \left(\frac{\pi}{2} + 2k\pi - \theta \right)  \sqrt{\frac  \beta {4\sqrt \beta - \alpha} },
\]
and for centers we get
\[
X^*(k) = d + \left(\frac{\pi}{2} + (2k+1)\pi - \theta \right)  \sqrt{\frac  \beta {4\sqrt \beta - \alpha}}.
\]

For the critical case (repeated $\lambda$) the potential function is 
\begin{equation}
U=-2 b^2 \alpha e^{\frac{-4 \sqrt{2}}{\sqrt{\alpha}} X} \left(X-d-\frac{\sqrt{2 \alpha}} {8}\right)
\label{criticalPotential}\end{equation}
and the center is given by
\[ X^* = d + \frac{\sqrt{2 \alpha}}{4}.\]

For the real case we have the potential function
\begin{equation}U = - b^2 a \sqrt{\alpha^2 -16 \beta} e^{-2aX} + 
 d^2 c  \sqrt{\alpha^2 -16 \beta} e^{-2cX},\
 \label{realPotential}\end{equation}
 and the center:
\[ \frac 1 {2(c-a)} \left( \log{\frac{c^2d^2}{a^2b^2}} \right),  \]
with $a$ and $c$ as given in Eq. (\ref{a_and_b}).

\section{Comparison of ODE and PDE models}

\begin{figure}[h!]
    \includegraphics[width=0.4\textwidth]{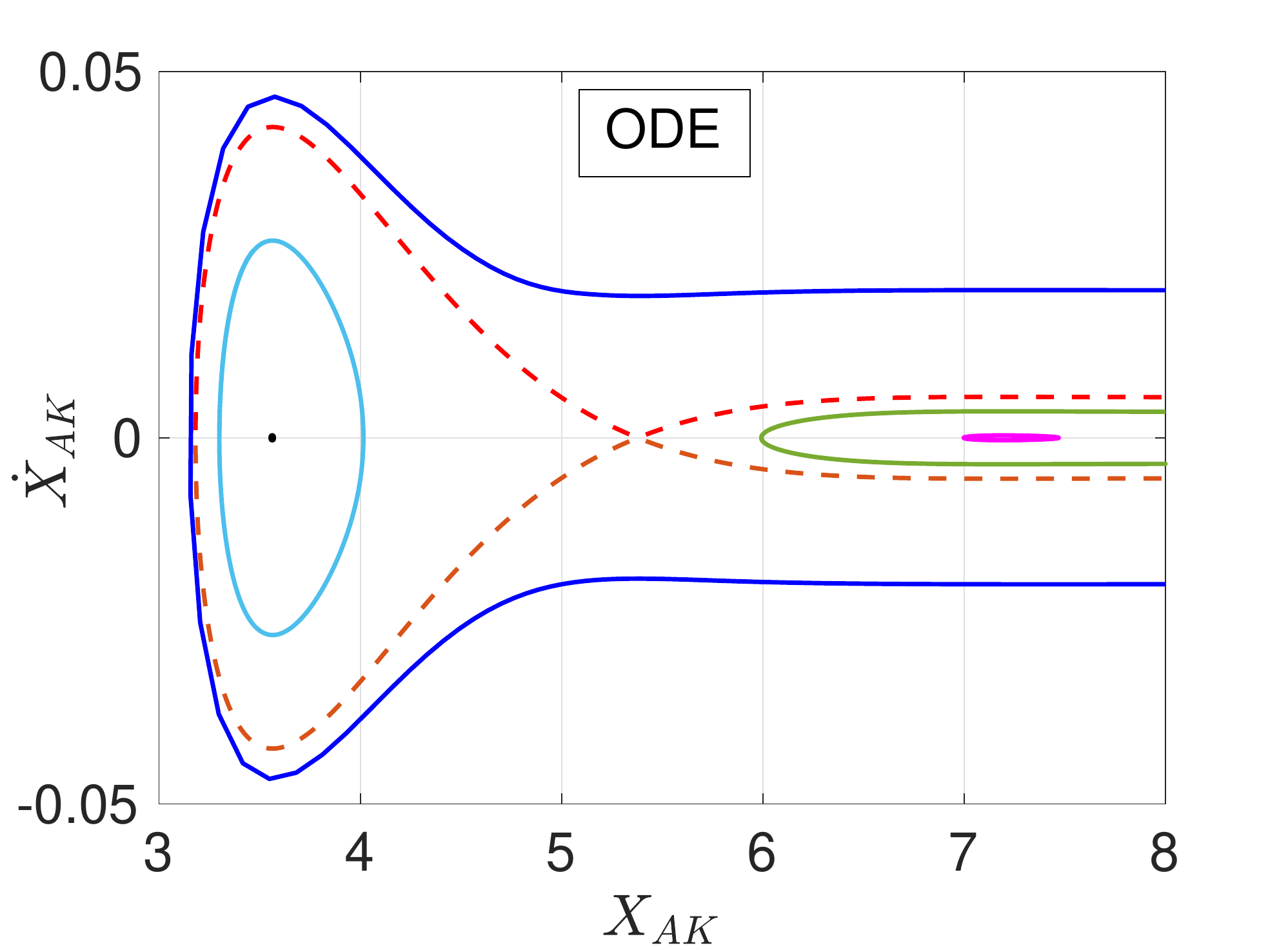}
       \includegraphics[width=0.4\textwidth]{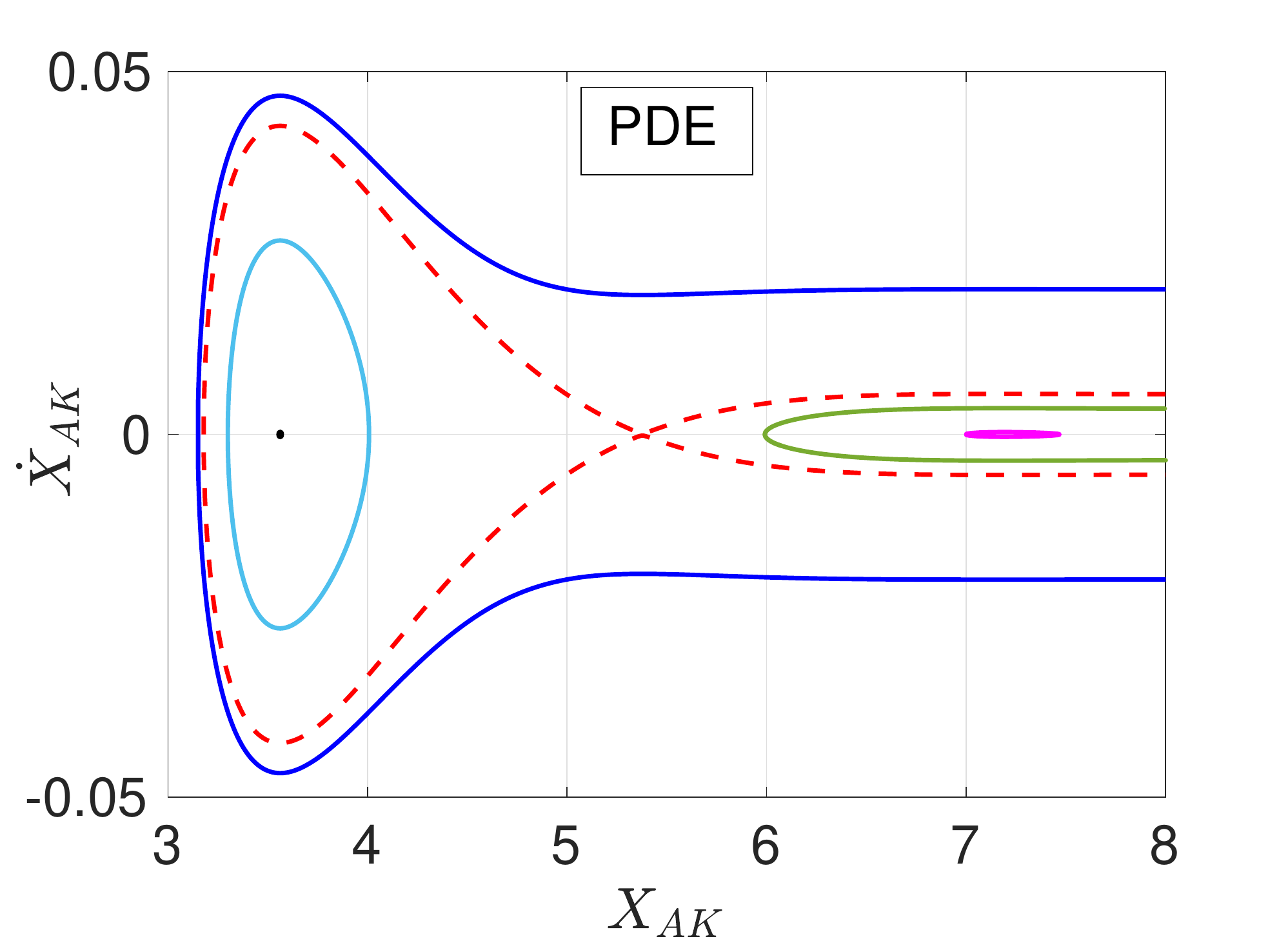}
    \caption{$\alpha =1$, $\beta = 1$, Phase portrait of 
    the ODE Equation (ODE) in comparison with Eq. (\ref{par-phi4}) (PDE). The blue solid curve corresponds to $X_{AK}(0)=8$, $\dot{X}_{AK}(0)=-0.02$.  The red dash-dotted curve: $X_{AK}(0)=8$, $\dot{X}_{AK}(0)=-0.00555$.  The light blue closed orbit: $X_{AK}(0)=3.3$, $\dot{X}_{AK}(0)=0$. The green curve corresponds to   $X_{AK}(0)=8$, $\dot{X}_{AK}(0)=-0.00356$. The pink solid closed orbit:  $X_{AK}(0)=7.4$, $\dot{X}_{AK}(0)=-0.0002$.  }
    \label{phaseplots}
\end{figure}

\begin{figure}[h!]
    \includegraphics[width=0.4\textwidth]{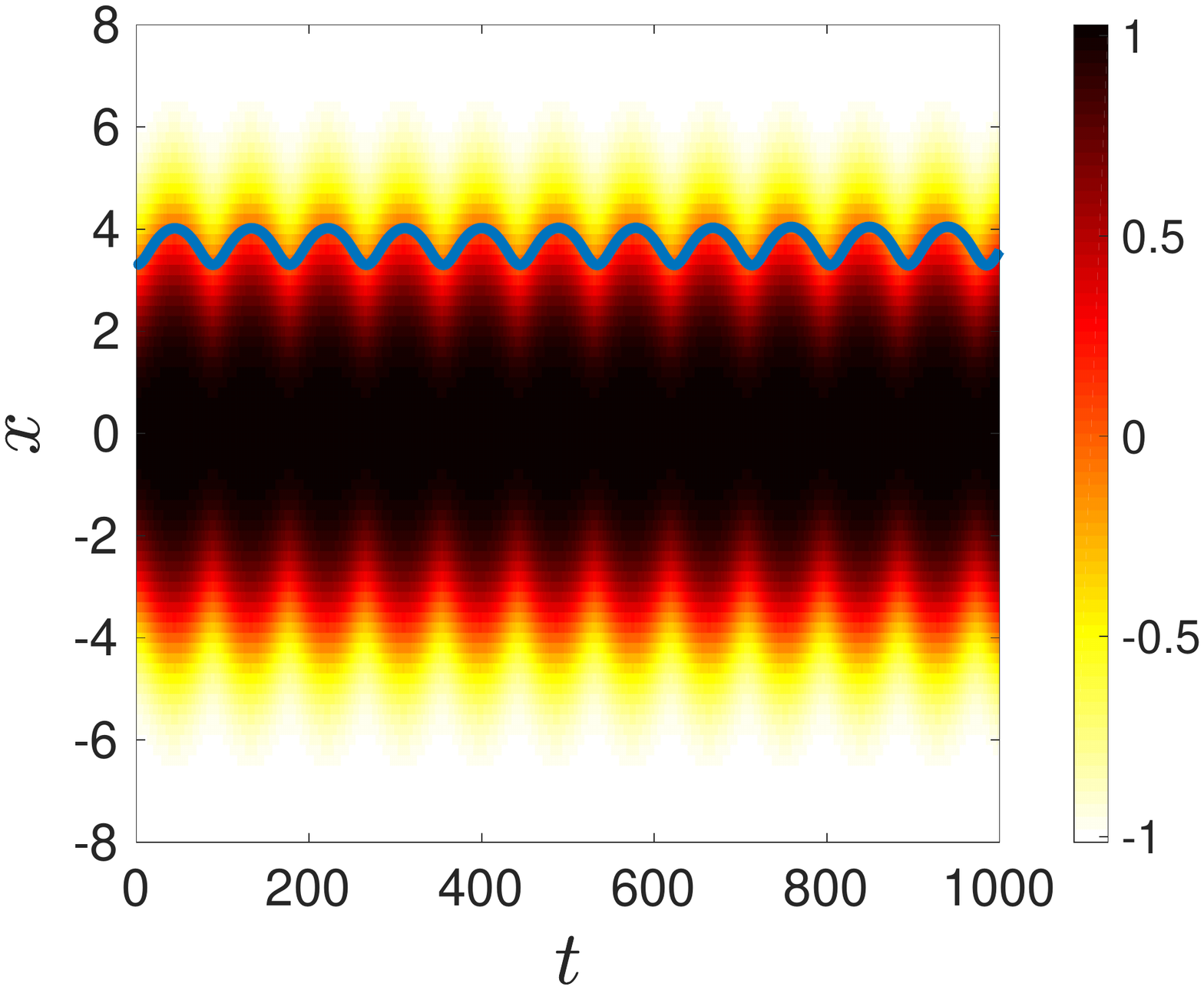}
       \includegraphics[width=0.4\textwidth]{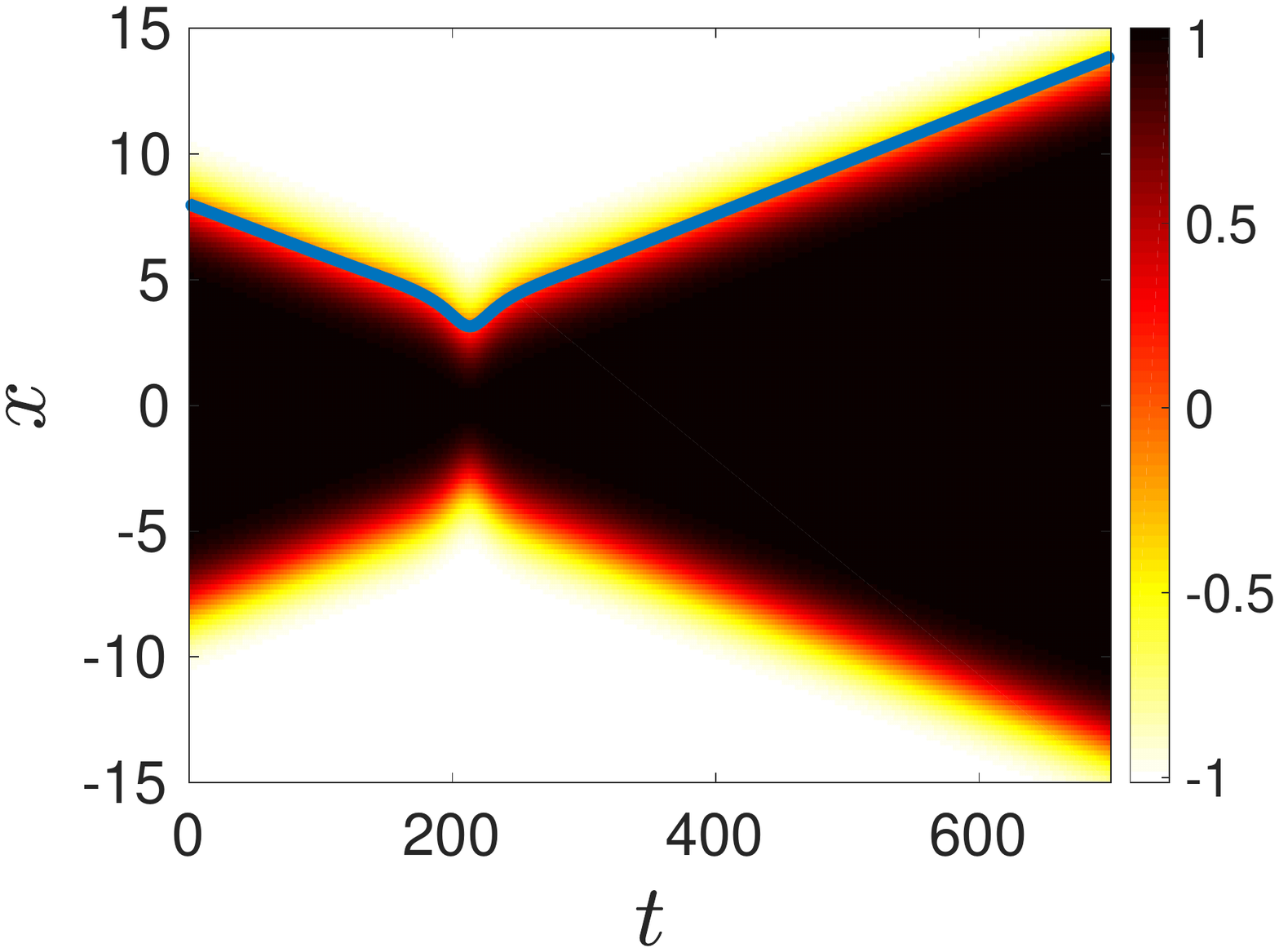}
    \caption{$\alpha =1$, $\beta = 1$. Comparisons of the PDE contour plot of the displacement field $u(x; t)$ and the ODE trajectory. (blue solid curve). Left: $x_0$ = 3.3, $v_{in} = 0$. This corresponds to the closed orbit (light blue) in Fig. \ref{phaseplots}. Right: $x_0$ = 8, $v_{in} = -0.02$. This corresponds to blue solid curve in Fig. \ref{phaseplots}.}
\label{contourPlots}\end{figure}

Using the expressions for the force we can now write an ODE for the time
evolution of the kink and antikink position for any given ($\alpha$, $\beta$) combination. In Table~\ref{accelTable} we have divided the force by the numerically calculated mass and used the curve-fitted values for $b$ and $d$ to get an acceleration expression for specific values of $\alpha$ and $\beta$. The corresponding ODE is then $\ddot{X}=-dU/dX$, where
the acceleration of the right hand side is
provided in Table~\ref{accelTable}. 
This ODE for the position of the one coherent structure (while the other
one is symmetrically located) is amenable to a phase portrait
analysis, as shown in Figure \ref{phaseplots} and a 
comparison 
with the corresponding PDE results of Eq.~(\ref{par-phi4})
can be obtained both at that level and at the 
spatio-temporal evolution one
as 
shown in Fig.~\ref{contourPlots}.

In Figure \ref{phaseplots} we show trajectories for the case $\alpha=1$ and $\beta=1$ (complex case) that illustrate behavior near the steady states of the PDE (the fixed points of the ODE). For these cases, there is clearly excellent agreement, between the ODE and PDE phase planes which validates our force calculations. The calculated force laws work well as long as the separation between kink and antikink is sufficiently large. 
In the case of the PDE, we identify the motion of the coherent structure by using the intersection of the kink or antikink with the horizontal axis as the position, and also find the corresponding speed, and
thus extract an effective phase portrait to be compared with the ODE results.
In Figure \ref{contourPlots} we show two of the trajectories from Figure \ref{phaseplots} as contour plots of the PDE with the ODE trajectory superimposed on top (in blue). 
The left one among them is a robust oscillation around a stable fixed point in the
form of a center (the light blue curve in Figure~\ref{contourPlots}).
The other is a trajectory that is scattered from the innermost potential
energy barrier --due to the presence of the 
innermost saddle point, corresponding to a maximum-- of the effective energy landscape and is thus reflected.
In this case, we see that the kinks do not make it to a collision but are rather
reflected due to their interaction landscape before the collision.

\begin{figure}[h!]
      \includegraphics[width=0.4\textwidth]{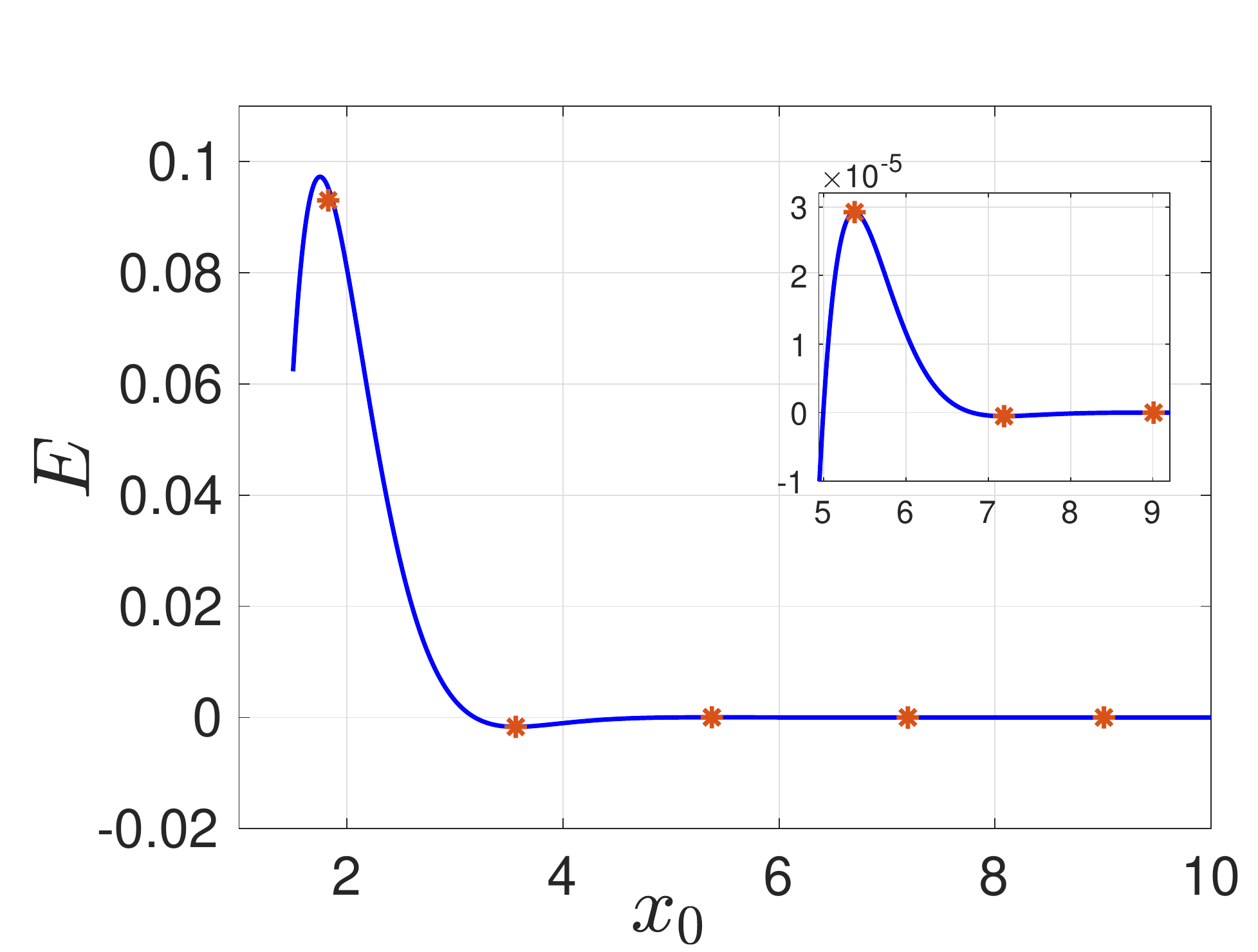}
      \includegraphics[width=0.4\textwidth]{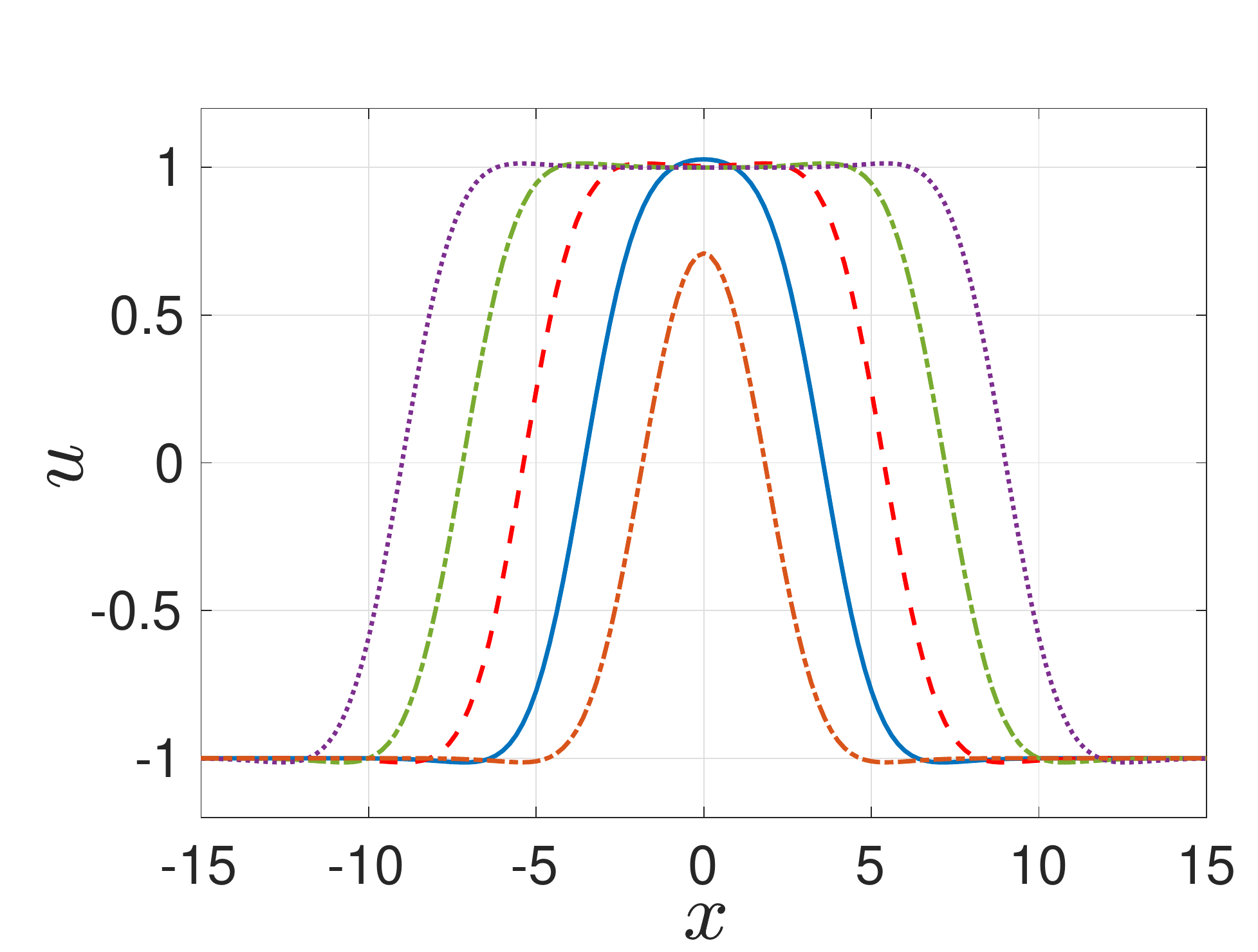}
    \caption{The left panel shows the energy vs $x_0$ for $\alpha =1$, $\beta = 1$. Blue curve is twice the potential function of the ODE for the complex case (given in Eq. (\ref{complexPotential})). The data points are the (normalized) potential energies of the steady states of the PDE at $x_0=1.825, 3.56, 5.38, 7.19, 9.01$ which are shown in the right panel. The need to multiply the potential function of the ODE by two when comparing ODE and PDE stems from
    the fact that the energy calculation using a steady state of the PDE involves two solitons - kink and antikink. The right panel presents the static, equilibrium solutions corresponding to $x_0 \approx 1.825$ (orange dashed-dot curve), $x_0 \approx 3.56$ (blue solid curve), $x_0 = 5.38$ (red  dashed curve), $x_0 \approx 7.19$ (green dashed-dot curve)  and $x_0\approx 9.01$
(purple dotted curve). Note that the steady state for the PDE occurs at $x_0 \approx 1.825$ but the fixed point of the ODE is at $x_0 \approx 1.75$.}
\label{energy-steady}
\end{figure}

For another perspective on the quantification of the 
agreement between PDE and ODE results, see Figure \ref{energy-steady}. In the
left panel of this figure, we show the potential energy plot of the ODE, again for $\alpha =1$, $\beta = 1$, in blue (using Eq. \ref{complexPotential}). The data points represent the potential energies of the steady states of the PDE calculated using the PDE energy $E=\int{\frac{1}{2}\alpha u_x^2+\frac{1}{2}\beta u_{xx}^2+\frac{1}{2} (u^2-1)^2 dx}$ of the associated steady state configurations. The corresponding steady-states themselves are shown in the same figure, right panel. Note that since the calculated potential energy curve of the ODE approaches zero as the separation distance increases, the potential energies of the steady states of the PDE must also be normalized (i.e., calibrated) so that the limiting value is zero (by subtracting the potential energy of a steady state with very large separation).  Again, clearly, the local maximum and minimum values of the ODE energy
landscape coincide with the potential energies of the corresponding steady states of the PDE. The local minima correspond to stable steady states of the PDE (centers of the ODE) and the local maxima correspond to unstable steady states of the PDE (saddle points of the ODE).
Importantly, aside from the center-most
potential energy maximum 
where the kink structures are so close that we cannot identify
them as independent entities (and thus we do not expect the collective 
coordinate characterization to be as accurate), we observe that the agreement
is very good. We remind the reader that the presence
of this oscillatory energy landscape, its
associated minima (centers) and maxima (saddles),
and the respective stationary PDE configurations
are distinctive features of the prevalence of the
biharmonic term and are genuinely absent in the 
harmonic case (and more generally for 
$\alpha>4 \sqrt{\beta}$, when the harmonic 
contribution is dominant).

\begin{figure}[h!]
 \includegraphics[width=0.4\textwidth]{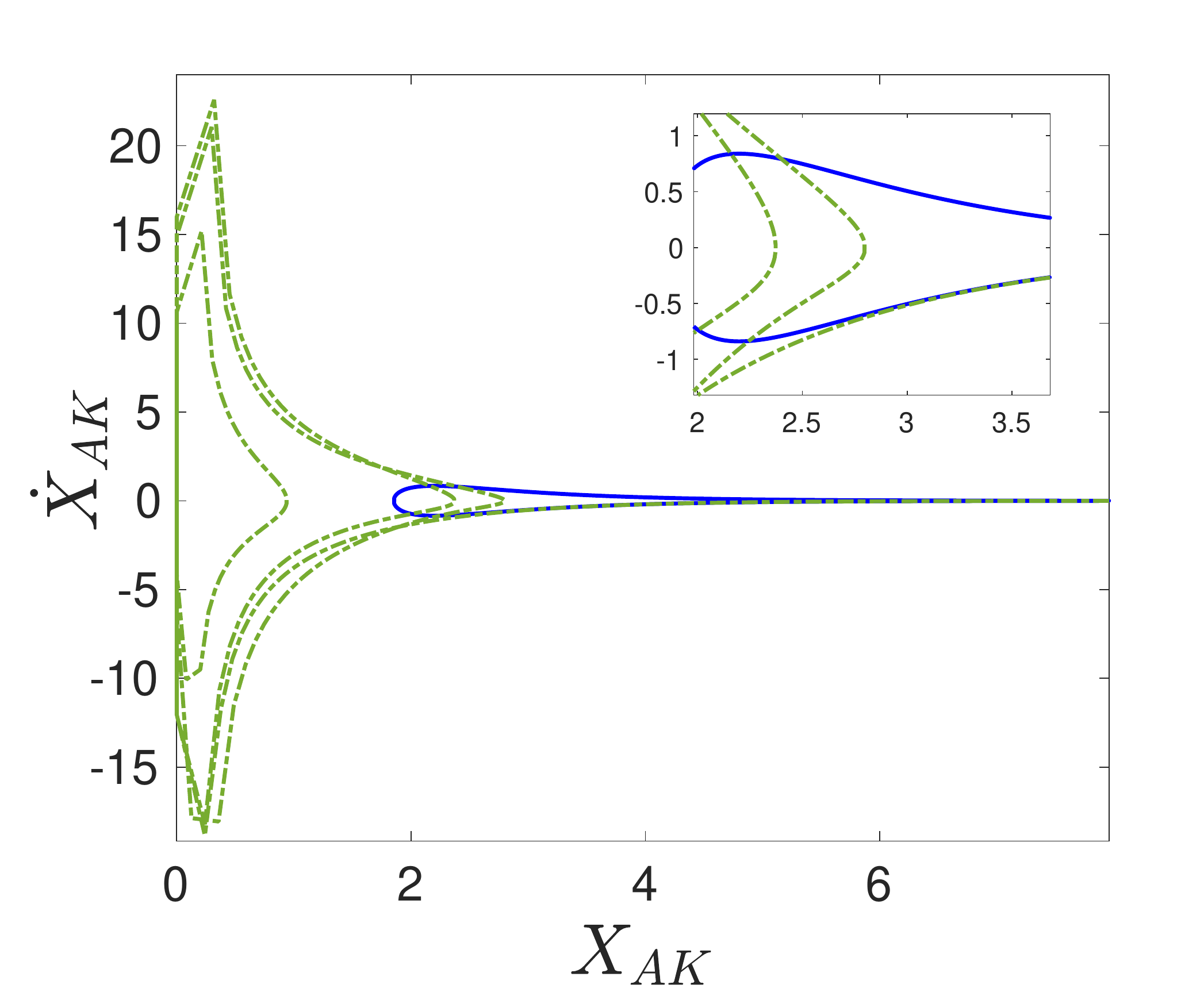}\hspace{0.5cm}
 \includegraphics[width=0.4\textwidth]{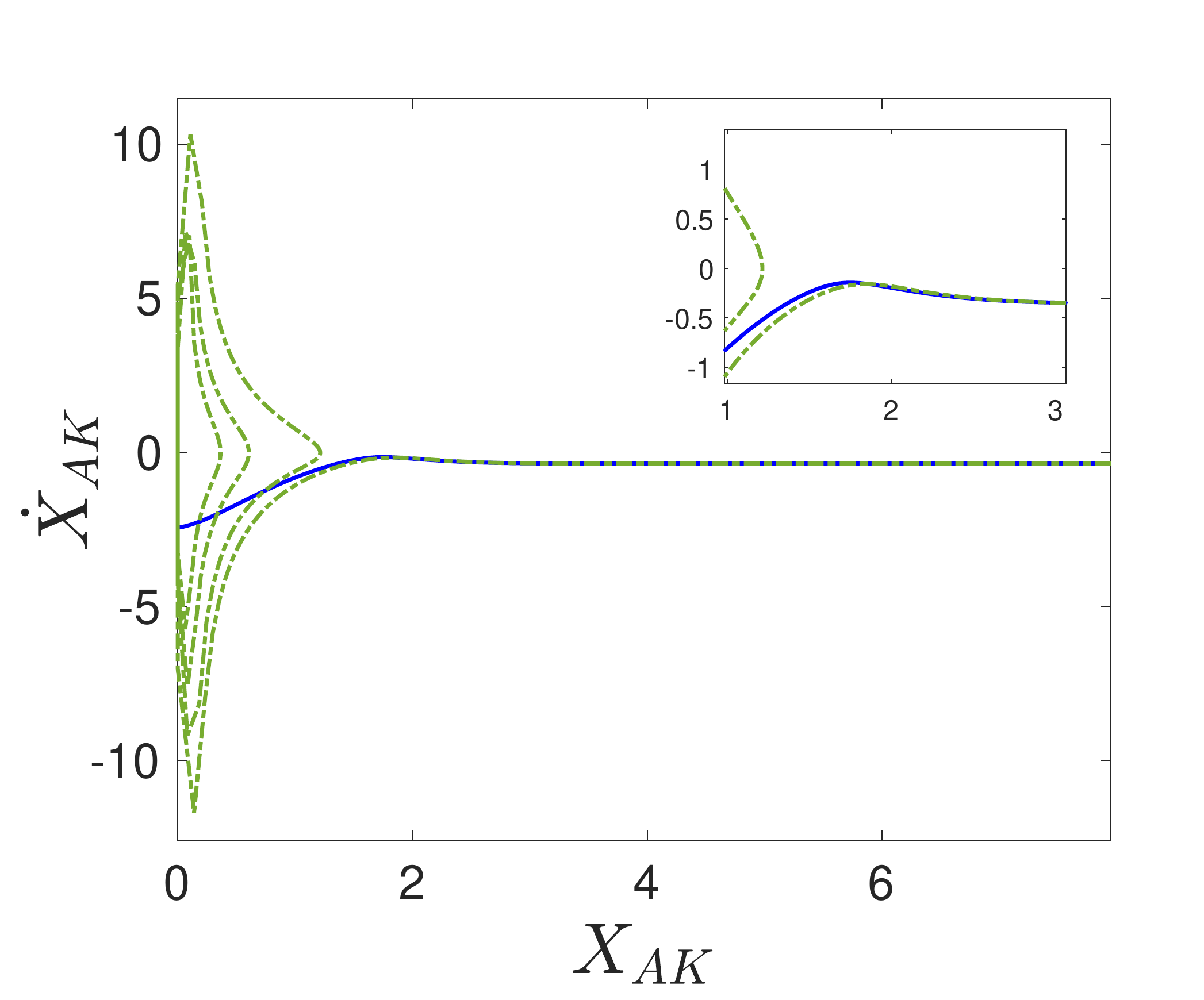}
 \caption{ The left panel shows phase plots for the real $\lambda$ case of $\alpha =5$ and $\beta = 1$ using initial conditions $X_{AK}(0)=8$ and $\dot{X}_{AK}(0)=-0.003593$. The right panel illustrates phase plots for the complex $\lambda$ case of $\alpha =1$ and $\beta = 1$ using initial conditions $X_{AK}(0)=8$ and $\dot{X}_{AK}(0)=-0.35$. In both cases,
 the ODE is shown by the blue solid curve and the PDE by the green dash-dotted curve. Insets show at what points the ODE model diverges from the PDE model. The ODE trajectory in the right panel is stopped at the point when $X_{AK}=0$ because it becomes physically unrealistic beyond that point.}
\label{PhaseDifference}
\end{figure}

While in Figure \ref{phaseplots} we showed example phase portraits that resulted in very proximal 
correspondence between PDE and ODE (for the complex case), in Figure \ref{PhaseDifference} we show phase trajectories for both the real and complex cases that illustrate at what point the PDE and ODE solution curves may depart from each other (recall that the dash-dotted green line represents the position of the antikink as measured by its intersection with the horizontal axis). 
In these cases, the kink and antikink get too close for the force law to remain valid. One can see that in the real case of the left
panel this occurs at about $X_{AK}=3$, while in the complex case 
of the right panel at about $X_{AK}=2$. Note also that the green curve (position of the antikink) indicates the formation of a bound state which is losing energy (in a way somewhat akin to a stable spiral but keeping in mind that in a bound state there are no longer an identifiable kink and antikink). Here, the important differences
of the PDE dynamics from the conservative ODE of 1 degree-of-freedom (dof)
become evident. The latter being energy conserving can only reflect
in some way, while the former can transfer energy from the kink translational
motion to other degrees of freedom (internal ones or radiation ones~\cite{ourp4}),
thus leading to the effective translational energy dispersion and thus
the apparent trapping of the kinks into a so-called bion state. 
Successive ``breathings'' of this bion state at the PDE level 
are mirrored in the
progressively inward green curves (carrying less and less energy).
At the ODE level, we make two more minor (in terms of the bigger
picture of our story), yet technically relevant observations. Given the
absence of ODE-PDE correspondence in the right panel we stop the
ODE evolution once the kink-antikink pair directly collides
(i.e., at $X=0$). On the other hand, the left panel has another
intriguing but non-physical trait: the double exponential force
(of opposite signs between the two exponentials) results in a landscape
with a local minimum very close to $X=0$. While this would result
in a turning point of $X\neq0$, we have found this feature to be
an artifact of the theory and its lack of accuracy in the 
immediate vicinity of $X=0$.
Let us reiterate, also in light of the above remarks,  that the ODE models are based only on the behavior of the tails of the kinks and antikinks. When a kink and antikink are involved in an interaction, the model makes sense only when the structures are well-separated. Therefore the ODE model should not be expected to reflect  the actual behavior of the system beyond the point where the waves are at a distance comparable to or smaller than their
width, at which time they essentially forego their individual character.

\begin{figure}[h!]
 \includegraphics[width=0.4\textwidth]{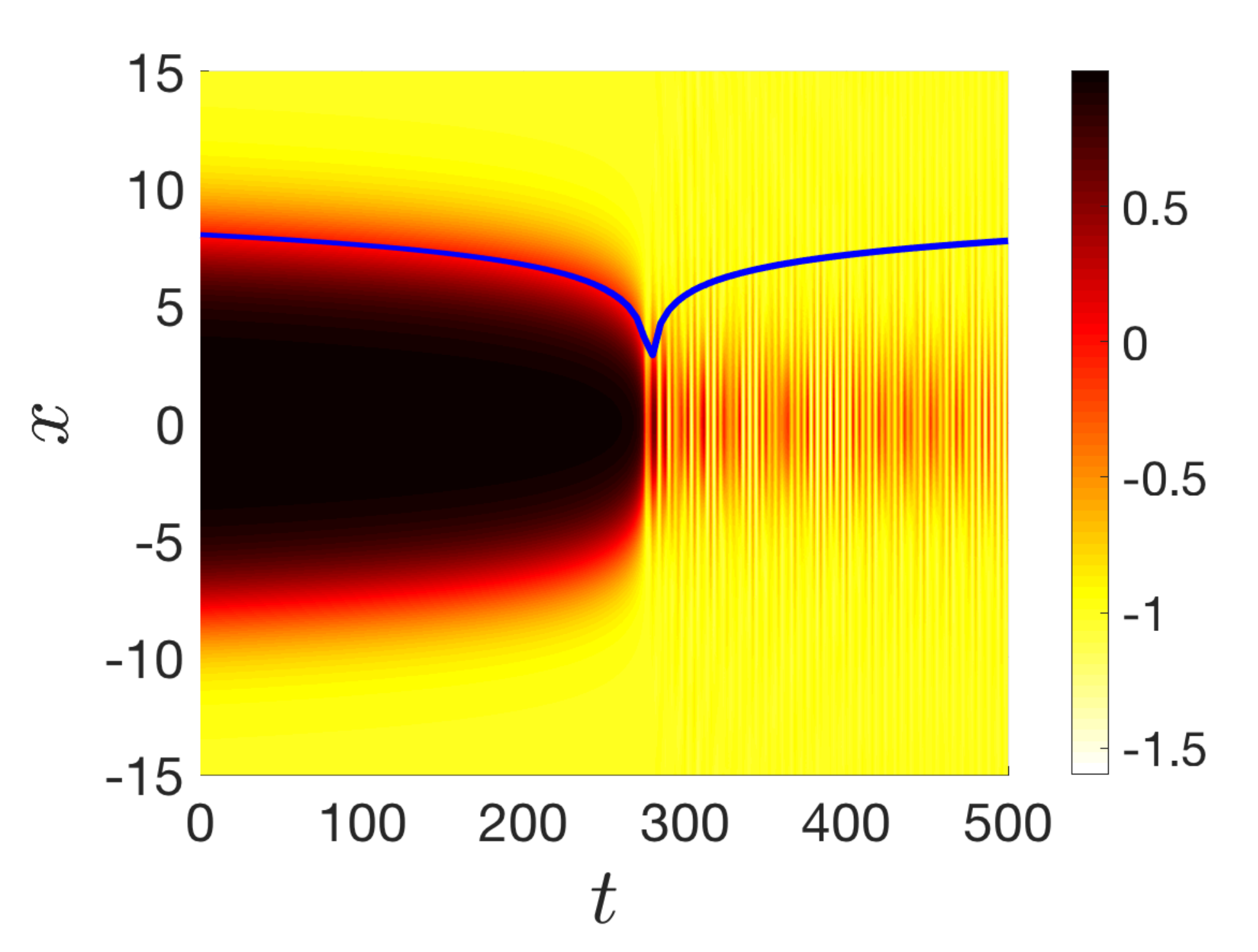}\hspace{0.5cm}
 \includegraphics[width=0.4\textwidth]{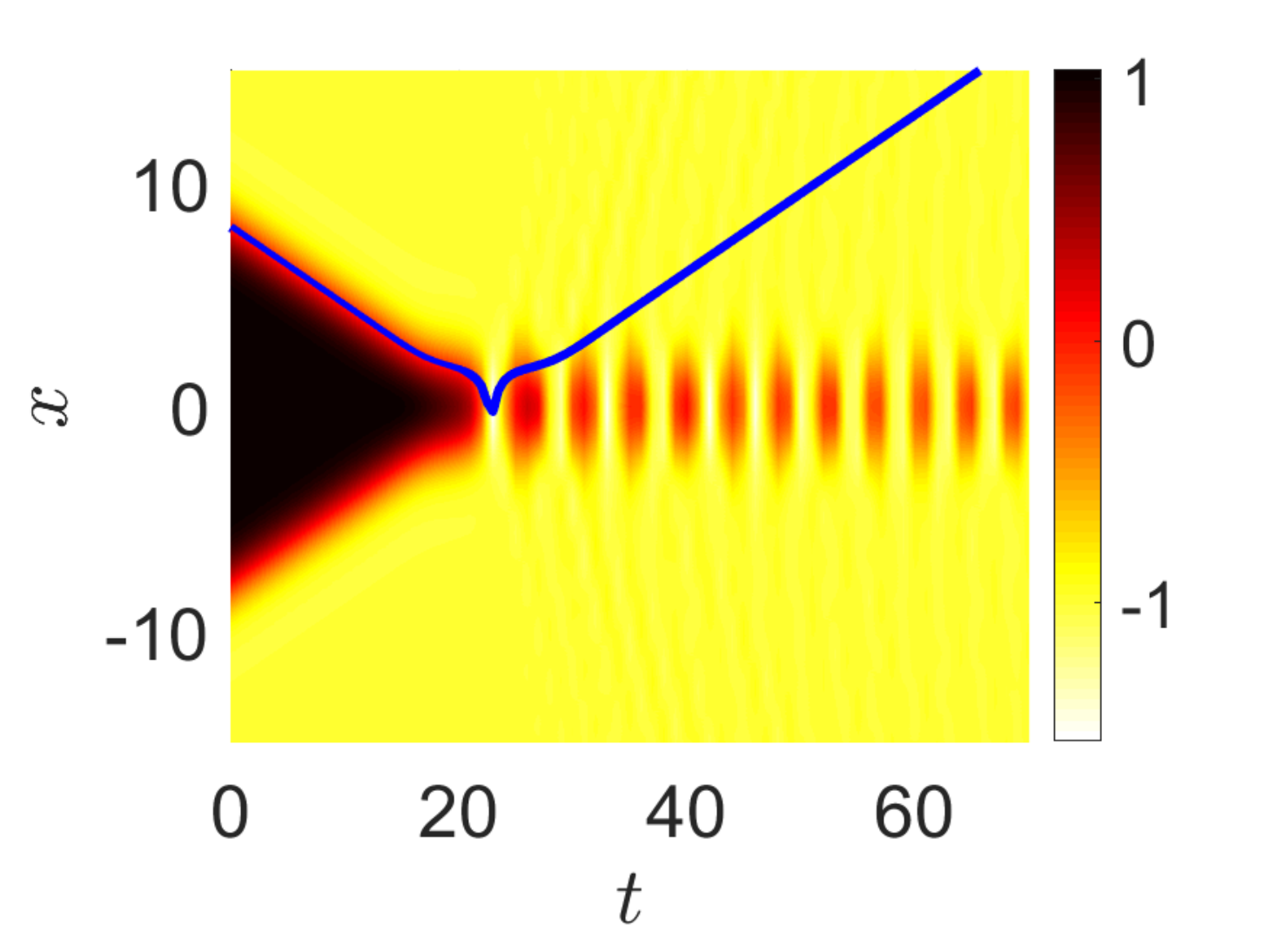}
 \caption{Contour plots of the PDE corresponding to the same parameter values and initial conditions as in the corresponding panels of Figure \ref{PhaseDifference}. The ODE trajectory is superimposed in blue.}
\label{badFitContours}
\end{figure}

In Figure \ref{badFitContours} we show contour plots for the same values of $\alpha$ and $\beta$ and the same initial conditions as in Figure \ref{PhaseDifference}, again with the ODE solution curves superimposed. 
I.e., these panels represent the spatio-temporal contour plot 
representation of the failure of the ODE theory to capture the
PDE dynamics, as explained in the above discussion.
As in Figure \ref{PhaseDifference} we can see that the ODE tracks the PDE simulation until around the time that the collision
occurs. As a result of the latter, at the PDE level, a bound state 
emerges, while
in the case of the ODE, the conservative nature of the 1 dof system does
not allow any scenario other than reflection.

\newpage
\color{black}
\section{Velocity in versus velocity out curves and Soliton Collisions}

We now investigate kink-antikink collisions in the context of escape velocity ($v_{out}$) and multi-bounce windows as a function of incoming velocity ($v_{in}$), 
in line with the extensive literature on the subject discussed in the 
Introduction (for a relatively recent summary in the
$\phi^4$ case, see, e.g.,~\cite{ourp4}).
For the latter model, it has been shown that there exists a critical $v_{in}$ value, which we label $v_{crit}$, such that for $v_{in}>v_{crit}$, the kink and antikink interact once and then separate forever. For $v_{in}<v_{crit}$ the kink and antikink can form a bound state, or can interact (bounce) any number of times, 
depending on $v_{in}$, before separating forever. Furthermore, 
it is well-established since the work of~\cite{Ann} for the
$\phi^4$ model, that
the bounce windows corresponding to different numbers of bounces are nested in a fractal pattern. For example, three bounce windows occur at the edges of two-bounce windows, four bounce windows occur at the edges of three-bounce windows, and so on. Also, the $v_{in} - v_{out}$ graph for a given window has the appearance of an inverted parabola, with the $v_{out}$ values going to zero at the edges of the window.

For a model with only a biharmonic term ($\alpha=0$ in this paper) it was shown in \cite{beam1} that two critical $v_{in}$ values exist, with $v_{1,crit}<v_{2,crit}$. $v_{2,crit}$ is similar to $v_{crit}$ for the $\phi^4$ model in that for $v_{in}>v_{2,crit}$ the kink and antikink interact once and then separate. For $v_{in}<v_{1,crit}$ the kink and antikink repel elastically before interacting. For $v_{1,crit}<v_{in}<v_{2,crit}$ the kink and antikink form a bound state. Near both critical values, we see oscillations in the $v_{in} - v_{out}$ graph, where the frequency of the oscillations rapidly increases as the critical values are approached.

In the model described in this paper, which is a mixture of the two cases just described, we find that by fixing $\beta$ at $\beta=1$ 
and letting $\alpha$ increase from zero to six, we see a transition from one case to the other. As $\alpha$ is increased, more and more bounce windows begin to populate the region between $v_{1,crit}$ and $v_{2,crit}$. At first these new bounce windows display oscillations in the $v_{in} - v_{out}$ graphs near  the edges of the windows, similar to what is seen in the bound-state region of the pure biharmonic case. With increasing $\alpha$ the oscillations diminish and the $v_{out}$ values at the edges of each window begin to approach zero as in the pure $\phi^4$ case. 
We will showcase these features qualitatively in the results that
follow. Nevertheless, the delicate nature of the associated computations renders
especially difficult 
the identification of effective ``critical points'' where the behavior
changes from the one reminiscent of the pure biharmonic problem
to that reminiscent of the pure harmonic one. 
In the case of the critical velocities, by rescaling we are  able to relate the solutions to Eq. (\ref{par-phi4}) for general $\alpha$ and $\beta=1$ to other combinations of $\alpha$ and $\beta$, as is now
shown.

Let \(u^{1,\beta}\) be a solution to 
\[u_{tt} = u_{xx} - \beta u_{xxxx}  + 2u -2u^3. \]
and consider the coordinate transformation 
\[ x \mapsto \xi = \frac x a.\]

In the new coordinate system the solution can be rewritten as \(u^{1,\beta}(x,t) = \tilde u(\xi,t)\). Of course, \(u^{1,\beta}_{tt} = \tilde u_{tt}\) and \(u^{1,\beta}_{x} = \frac 1a \tilde u_{\xi}\), therefore \(\tilde u\) obeys the equation
\[\tilde u_{tt} = \frac 1 {a^2} \tilde u_{\xi\xi} - \frac \beta {a^4} \tilde u_{\xi\xi\xi\xi}  + 2 \tilde u -2 \tilde u^3. \]

For \(a ^4 = \beta\) we get
\[\tilde u_{tt} = \frac 1 {\sqrt \beta} \tilde u_{\xi\xi} -  \tilde u_{\xi\xi\xi\xi}  + 2 \tilde u -2 \tilde u^3 \]
so, \(\tilde u(\xi,t) \equiv u^{\alpha,1}(\xi,t)\)
or, 
\[    u^{1,\beta}(x,t) = u^{\alpha,1} \left( \frac x {\beta^{1/4}} \right)\]
for \(\alpha = \frac 1 {\sqrt \beta} \).

Therefore we can obtain solutions to the model for the parameters \(\alpha = 1\) and \(\beta\), using the solution for  parameters \(\alpha \) and \(\beta = 1\).
Then, using this coordinate transformation, we find that the critical velocity \(v_{1,crit}^{1,\beta} \) of the solution \(u^{1,\beta}\) can be expressed in terms of the corresponding critical velocity \(v_{1,crit}^{\alpha,1} \) of the \(u^{\alpha,1}\) solution as

\begin{equation}
    v_{1,crit}^{1,\beta} = \frac {dx}{dt} =a \frac {d\xi}{dt} = \beta^{1/4} v_{1,crit}^{\alpha,1}, 
    \label{alphaBetaScaling}
\end{equation}
where \(\alpha = \frac 1 {\sqrt \beta} \).

Notice that when \(\beta\) becomes large enough, we get \(v_{1,crit}^{\alpha,1} \approx v_{1,crit}^{0,1}\), so
\[v_{1,crit}^{1,\beta} \sim  \beta^{1/4} v_{1,crit}^{0,1}. \]
Similarly
\begin{equation*}\label{v2alpha}
v_{1,crit}^{\alpha,1} \sim  \sqrt \alpha v_{1,crit}^{1,0}. 
\end{equation*}
Furthermore, the above equations hold when $v_{1,crit}^{\alpha,1}$ is replaced by $v_{2,crit}^{\alpha,1}$.

\begin{figure}[h!]
    \includegraphics[width=0.4\textwidth]{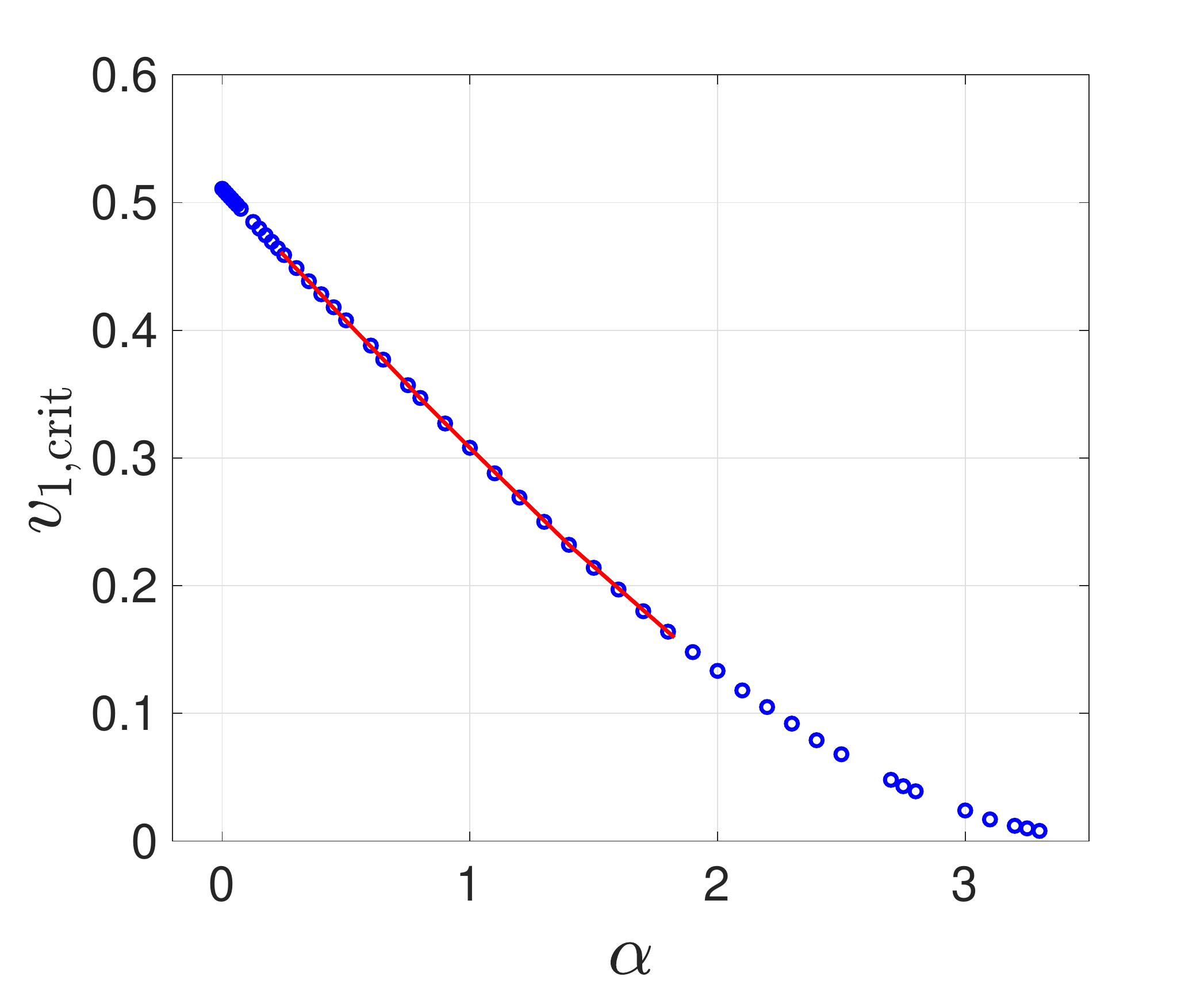}
    \includegraphics[width=0.4\textwidth]{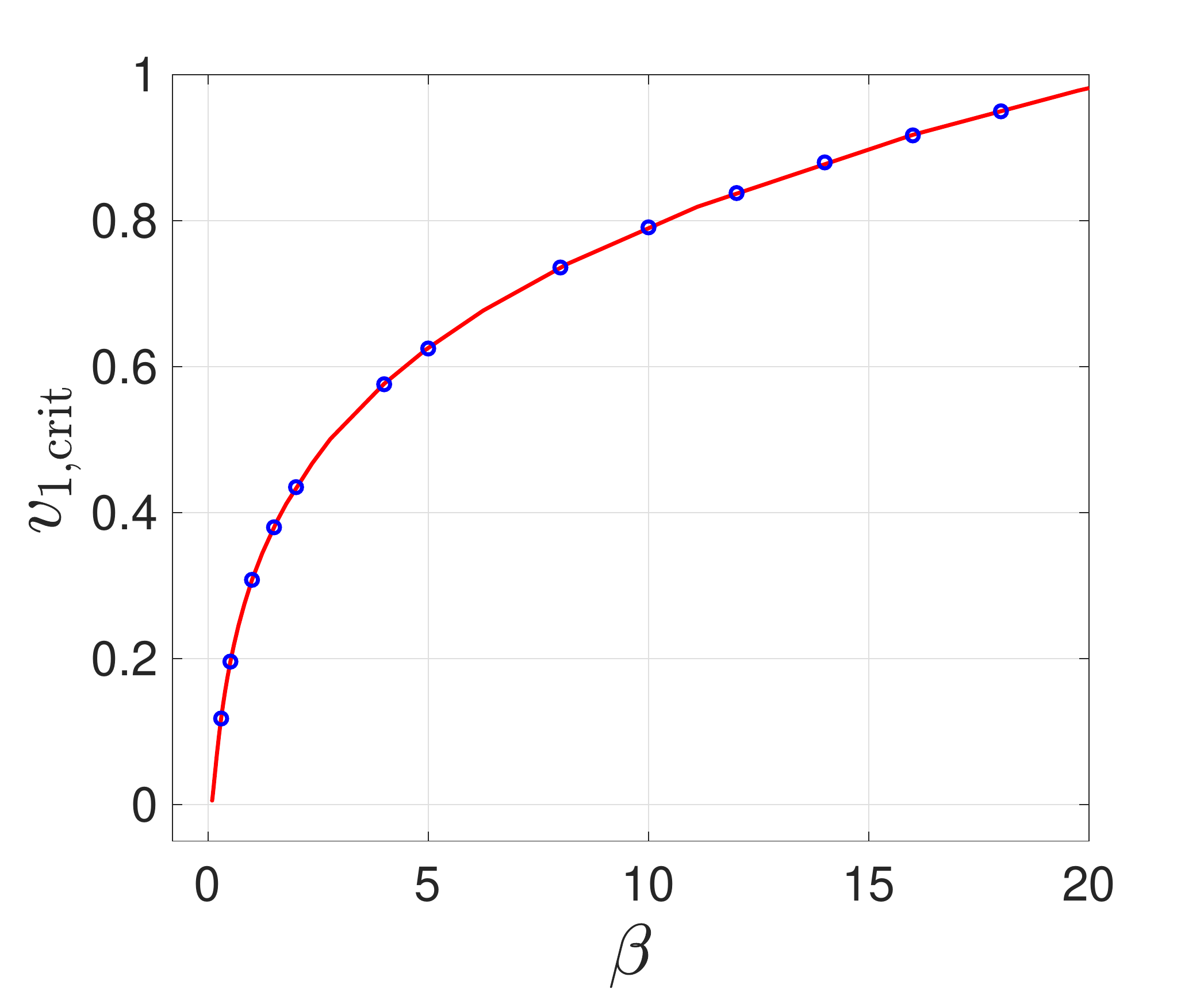}
    \caption{The blue circles on both panels are obtained by the numerical simulation of Eq. (\ref{par-phi4}) where left panel represents  $v_{1,crit}$ vs $\alpha$ when $\beta=1$ and the right panel 
    represents $v_{1,crit}$ vs $\beta$ when $\alpha=1$. The red solid curve on the left panel is obtained by applying the formula $v_{1,crit}^{\alpha,1}=\sqrt{\alpha} v_{1,crit}^{1,\beta}$ where $\beta=\frac{1}{\alpha^2}$ to the numerically obtained data (blue circles) on the right. The red solid curve on the right panel is obtained by applying the formula $v_{1,crit}^{1,\beta}=\beta^{1/4}v_{1,crit}^{\alpha,1}$ to the numerically obtained data (blue circles) on the left. }
    \label{vcrit1}
\end{figure}

\begin{figure}[h!]
    \includegraphics[width=0.4\textwidth]{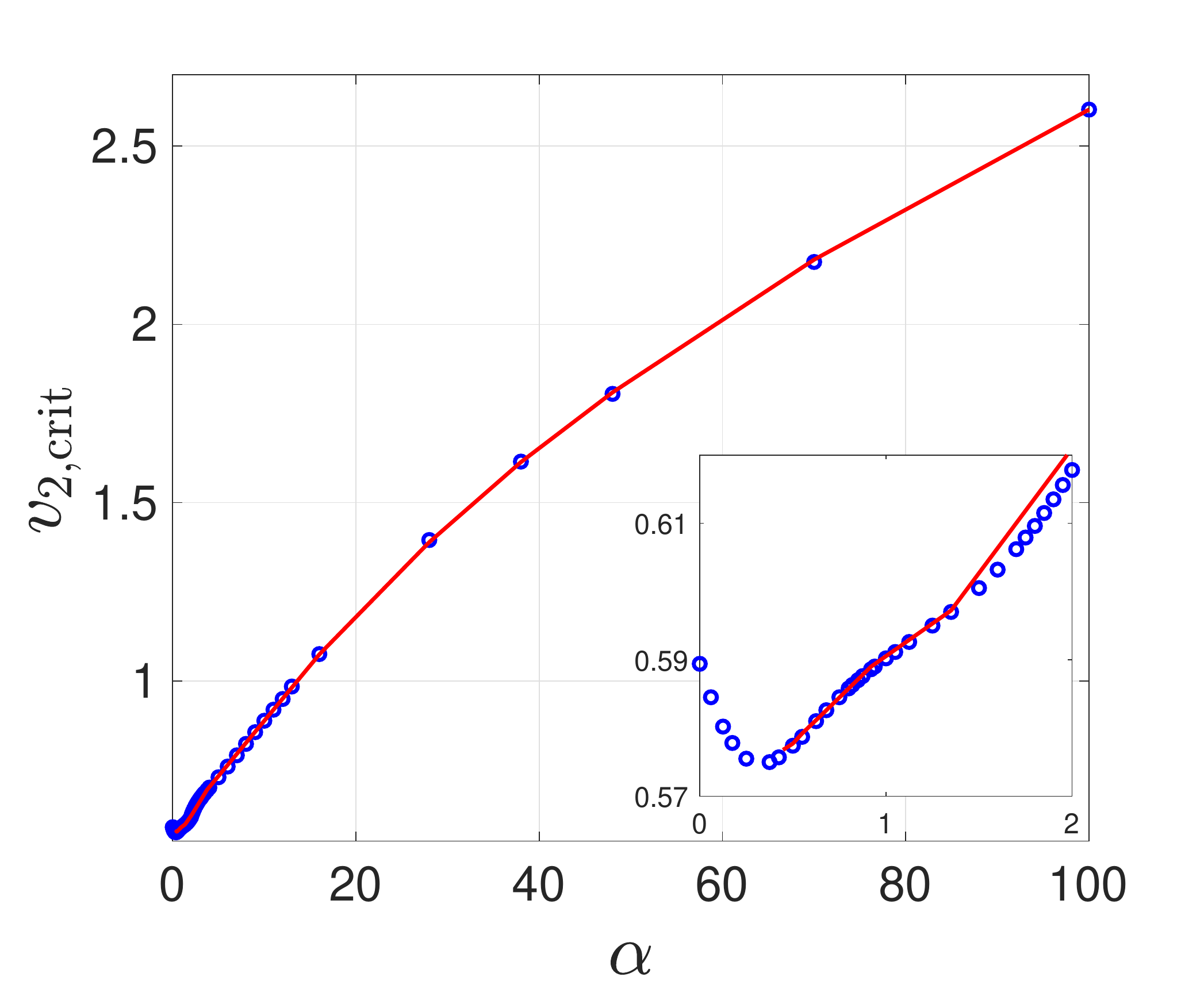}
    \includegraphics[width=0.4\textwidth]{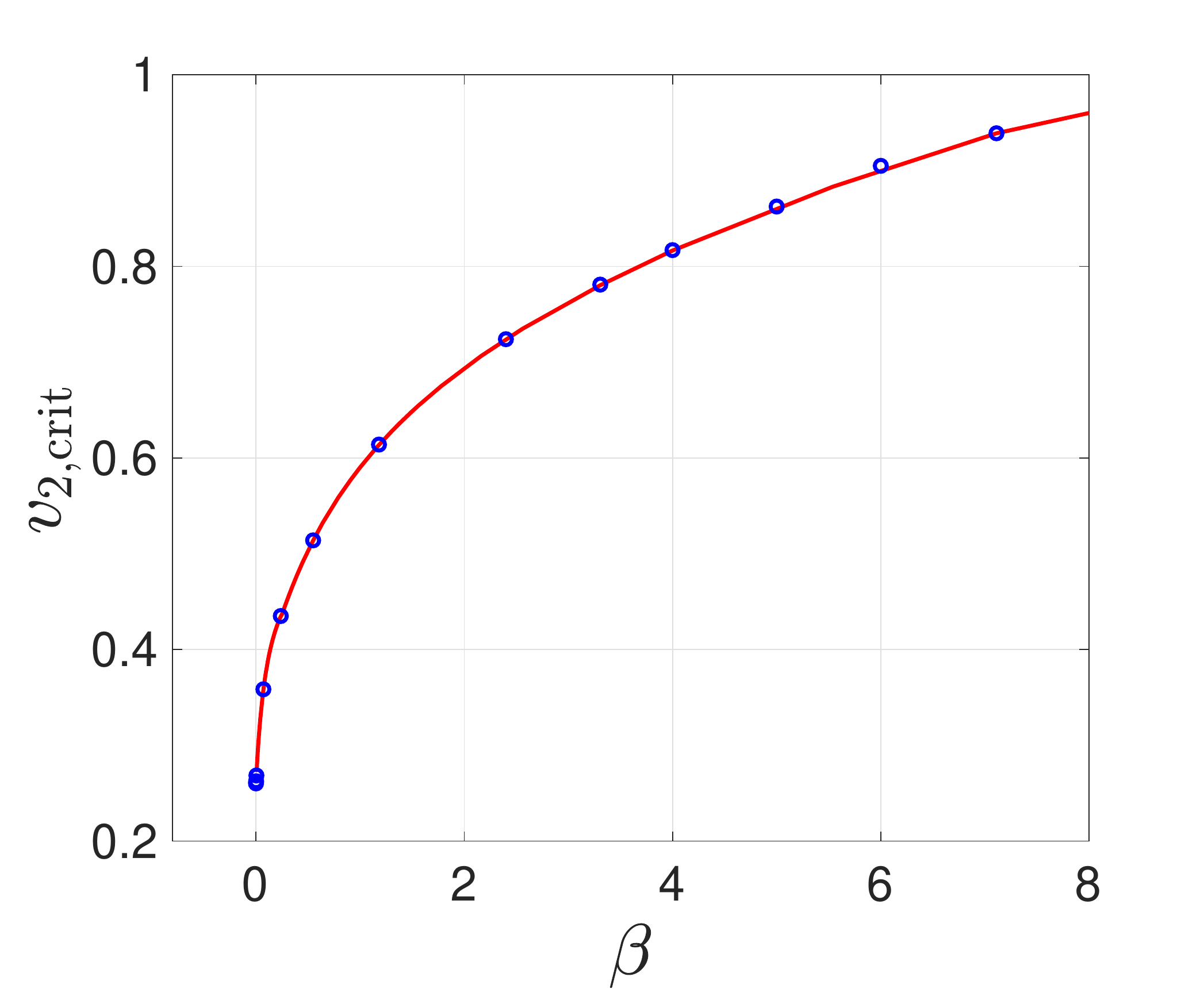}
    \caption{The left panel shows $v_{2,crit}$ vs $\alpha$ when $\beta=1$ and the right panel shows $v_{2,crit}$ vs $\beta$ when $\alpha=1$.  The blue circles and the red solid curves were obtained as described in Figure \ref{vcrit1}.}
    \label{vcrit}.

\end{figure}

In Figure \ref{vcrit1} we show graphs of $v_{1,crit}$ versus $\alpha$ for $\beta=1$ (left panel) and $v_{1,crit}$ versus $\beta$ for $\alpha=1$ (right panel). Figure \ref{vcrit} is similar, but for $v_{2,crit}$. The blue circles on all panels are obtained by the numerical simulation of Eq. (\ref{par-phi4}) where the left panels represent  $v_{1,crit}$ vs $\alpha$ when $\beta=1$ and the right panels represent $v_{1,crit}$ vs $\beta$ when $\alpha=1$. In the panels of both figures, the red curves are obtained from the transformation given by Eq. (\ref{alphaBetaScaling})  (plotted without markers for the transformed points and with connecting lines in order to make the graph more readable). 
The red curves are included to demonstrate the validity of Eq. (\ref{alphaBetaScaling}) in comparison with direct PDE simulations.

\begin{figure}[h!]
    \includegraphics[width=0.4\textwidth]{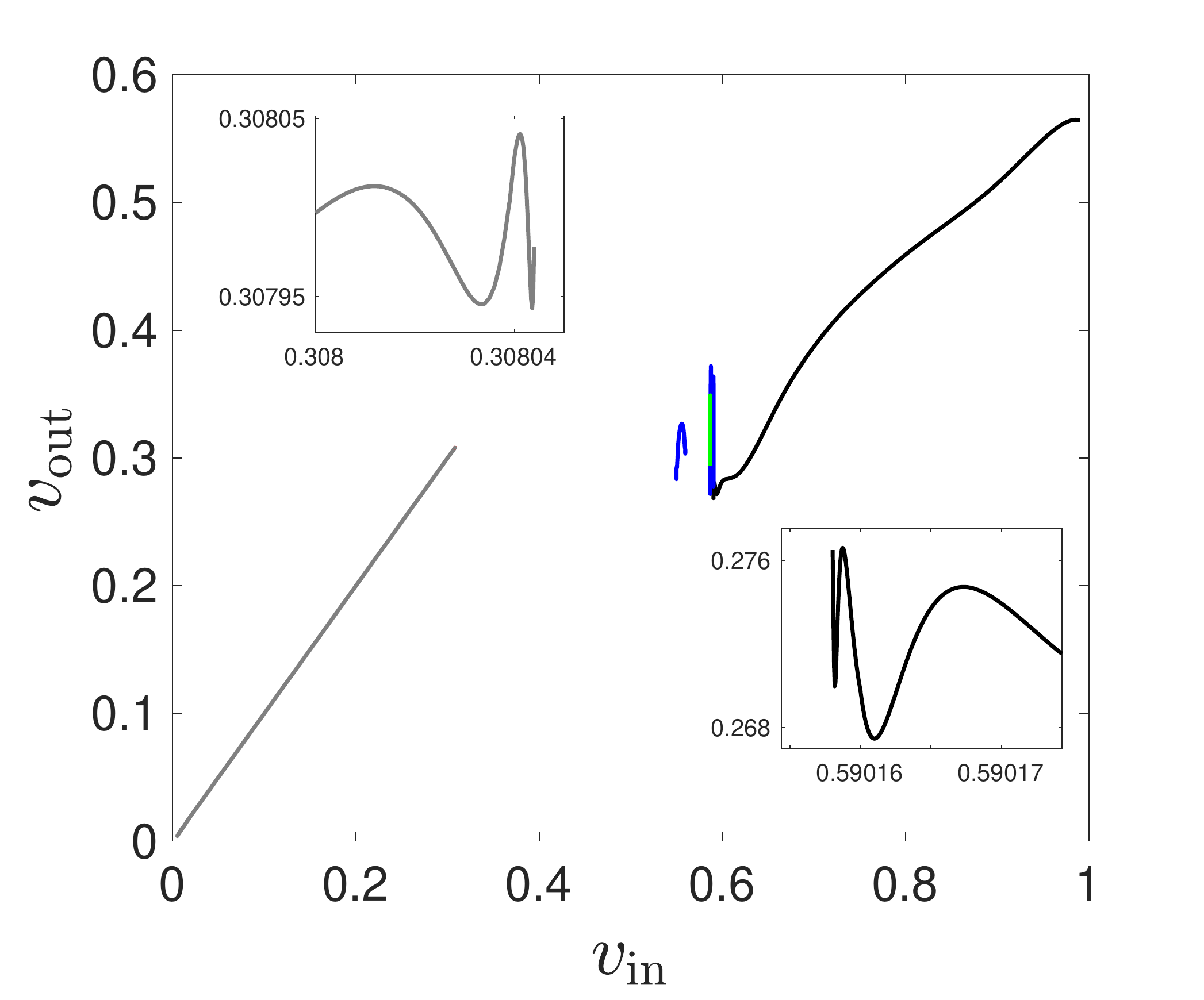}
    \includegraphics[width=0.4\textwidth]{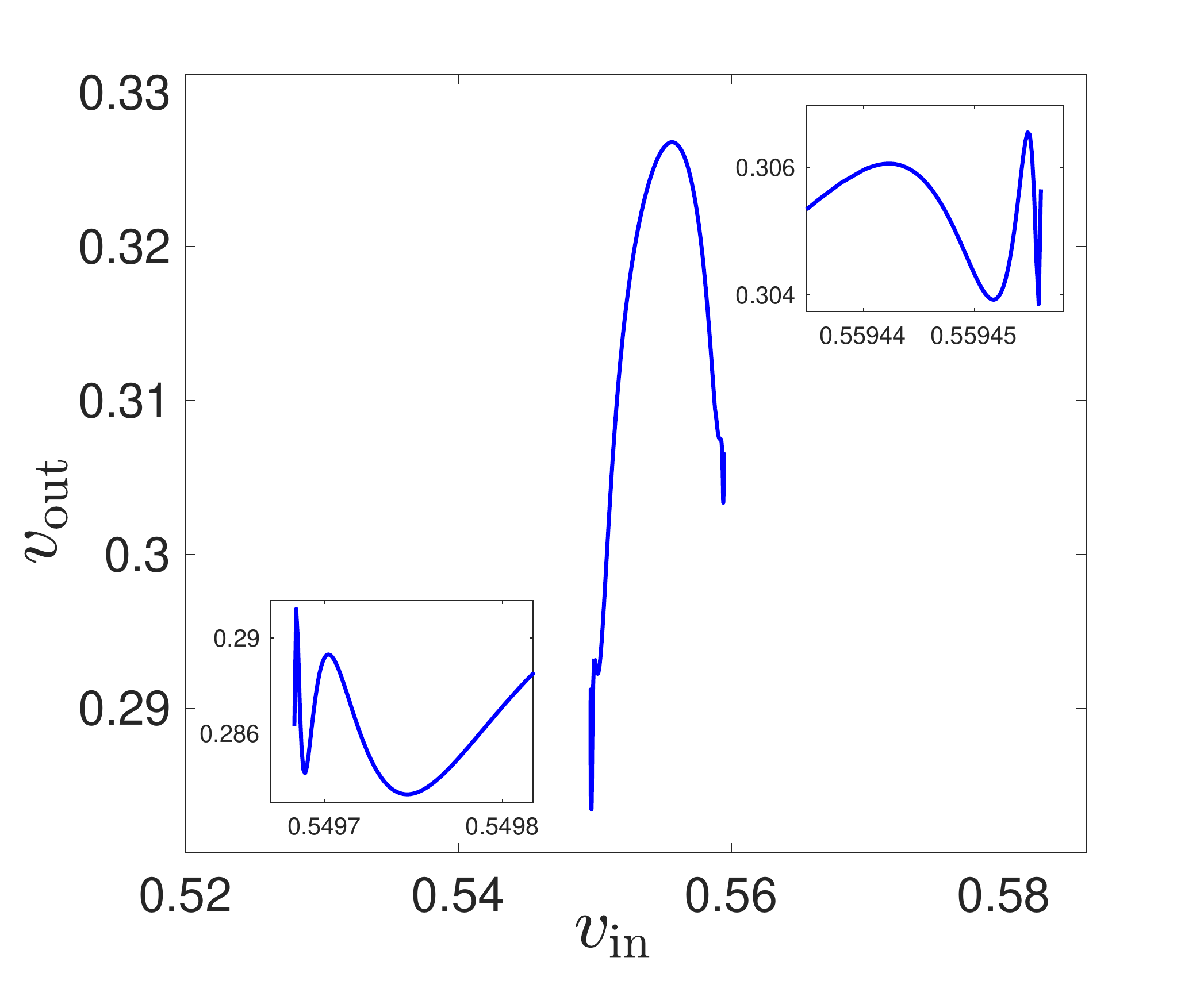}
    \includegraphics[width=0.4\textwidth]{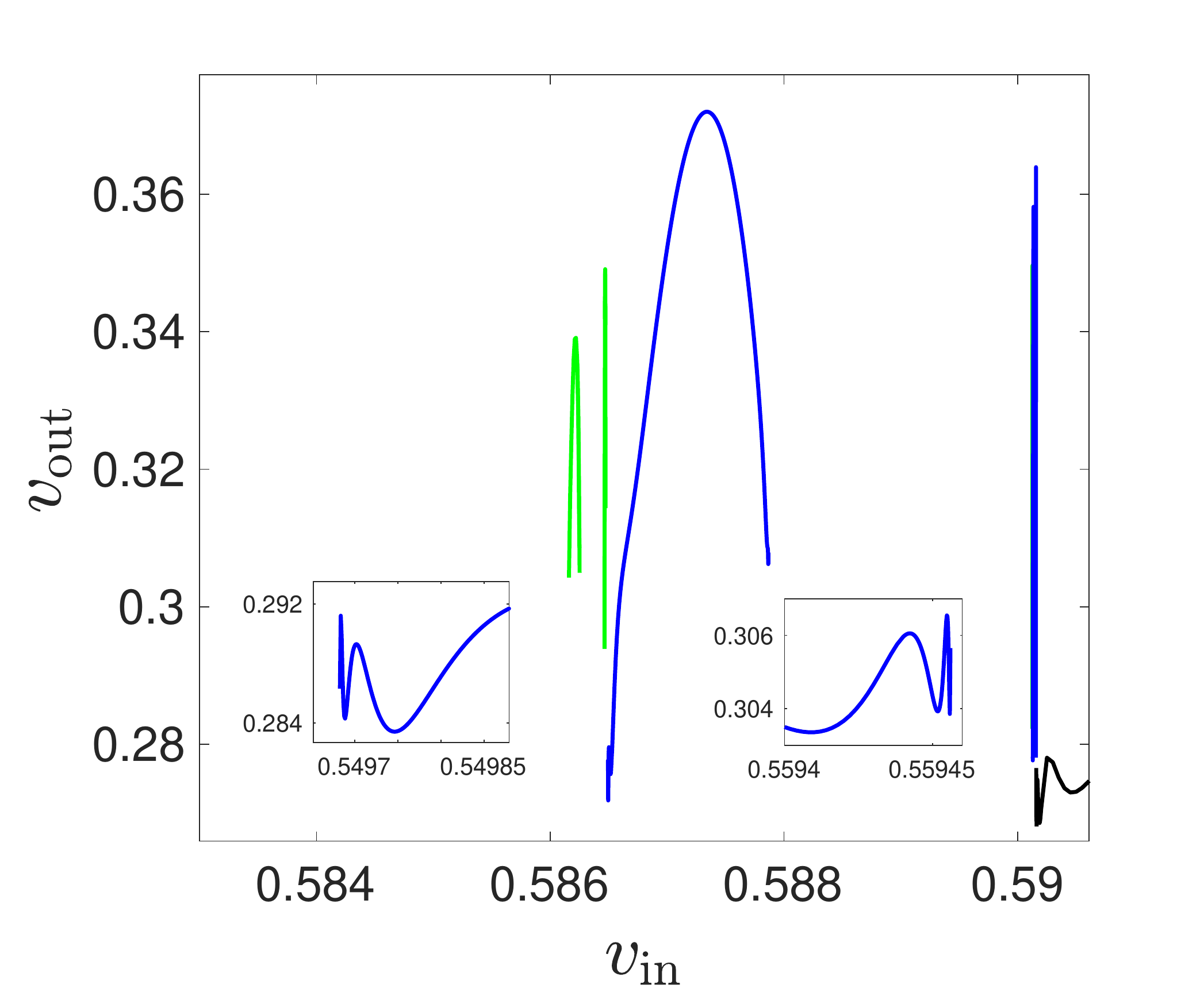}
     \includegraphics[width=0.4\textwidth]{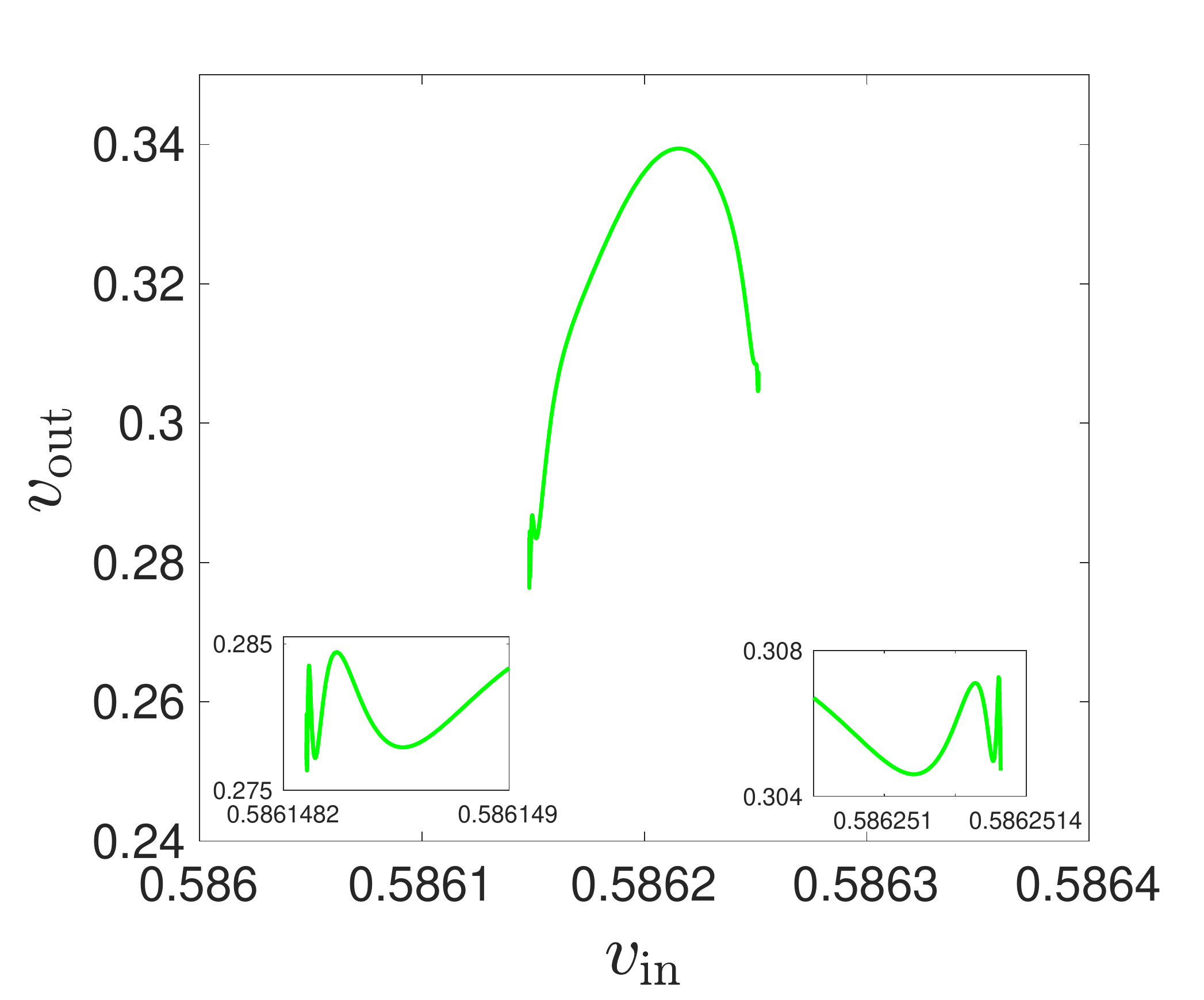}
    \caption{The top left panel shows $v_{\rm out}$ vs $v_{\rm in}$ when $\alpha=1$ and $\beta=1$ with $v_{2,crit} \approx 0.5902$. The top right panel is the zoom-in about the first two-bounce curve. The bottom left panel is the zoom-in about the two three-bounce windows right before the critical velocity $v_{2,crit}$. 
    The bottom right panel is the zoom in about the  
    leftmost three-bounce window on the bottom left panel. In both top right and bottom panels, the tails and their oscillatory behaviors are shown. One-bounce windows in the figures are in solid black. Two-bounce windows are in blue and three bounce windows are in green. The gray solid line on the top left panel is when the kink-antikink repel each other elastically. }
    \label{vin_vout}
\end{figure}

Having identified the critical point scaling relations, we now turn
to a direct examination of the collision features and associated multi-bounce
windows.
In Figure \ref{vin_vout} we show $v_{in}-v_{out}$ curves for $\alpha=1$ and $\beta=1$. For fixed $\beta=1$ we know that the force law changes from the complex $\lambda$ case to the real $\lambda$ case at $\alpha=4$, so we expect the case $\alpha=1$ to be somewhat similar to the pure biharmonic case of $\alpha=0$. In the upper left panel of Figure \ref{vin_vout} we see that the elastic collision region corresponds to $0<v_{in}<v_{1,crit} \approx 0.30805$ and the one-bounce region corresponds to $v_{2,crit} \approx 0.5902<v_{in}<1$. Note that we have chosen not to include $v_{in}$ values greater than one. The region $v_{1,crit}<v_{in}<v_{2,crit}$, which corresponds to a bound state when $\alpha=0$, is beginning to be populated by two and three bounce windows, a byproduct
of the inclusion of the quadratic dispersion. 
The top right panel shows the first two-bounce window. The bottom left panel shows the next two-bounce window, with three-bounce windows appearing just to the left. The bottom right panel shows the first three-bounce window in more detail. All windows display the characteristic oscillations at the edges. Notice the important features of this case: on the one hand, the
multi-bounce windows (which did {\it not} appear in the pure biharmonic
case) are now present. On the other hand, they do not terminate as, e.g., 
in the case of the standard $\phi^4$ model~\cite{ourp4,lizunova}, 
but rather have the oscillatory
terminations (with progressively shorter periodicity) encountered previously
in~\cite{beam1} for the pure biharmonic case, Moreover, we have encountered
a feature also absent in the standard (pure) $\phi^4$
case, namely higher-bounce
windows appear only on one side (to the left) of the two bounce windows,
while it is well-known~\cite{Ann} that they appear on both sides in the
pure harmonic $\phi^4$ problem.

\begin{figure}[h!]
   
    \includegraphics[width=0.4\textwidth]{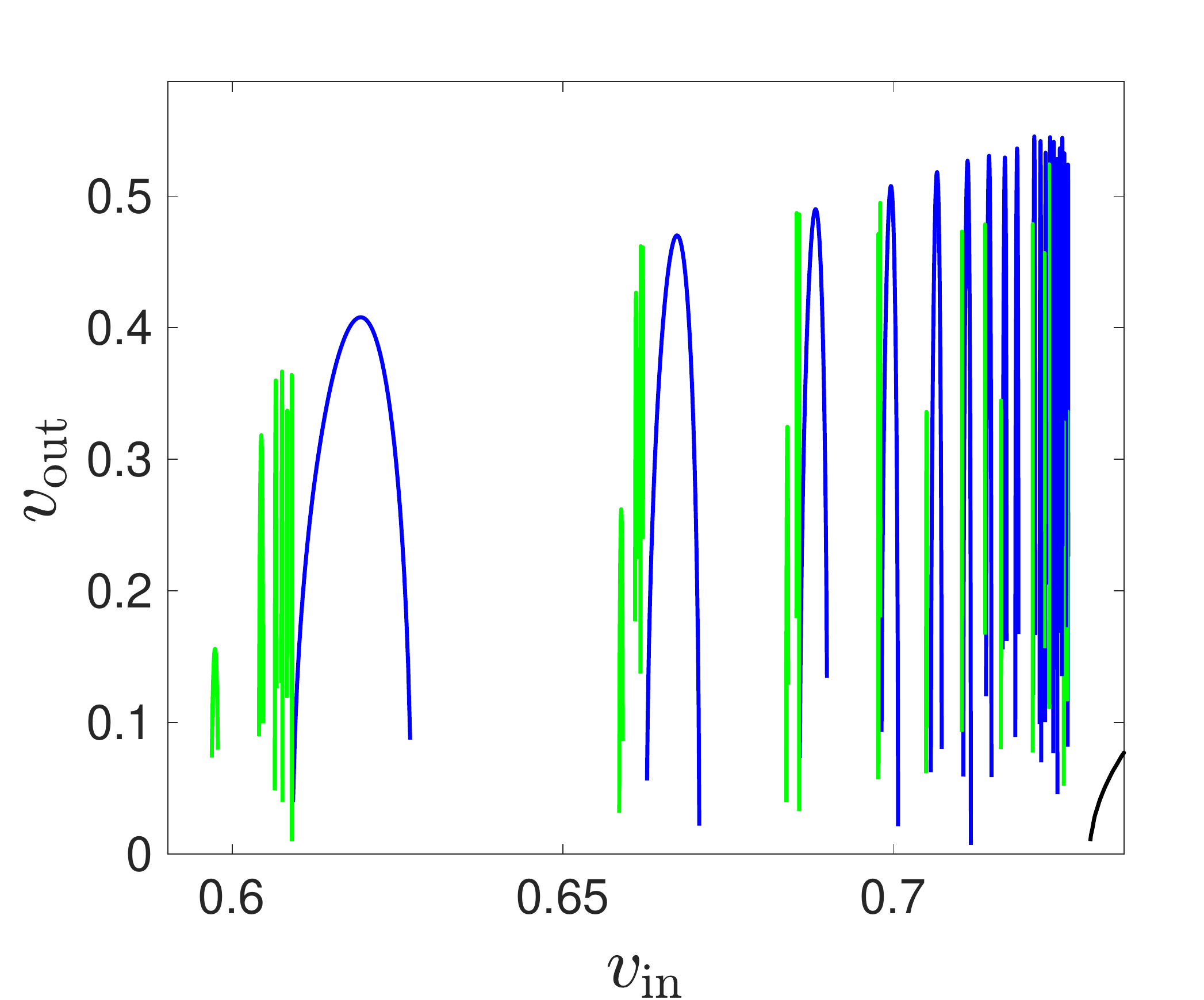}
    \caption{$v_{\mathrm{out}}$ vs $v_{\mathrm{in}}$ when $\alpha=5$ and $\beta=1$ with $v_c \approx 0.7295$. 
    The one-bounce window is in solid black. Two-bounce windows are in blue and three-bounce windows are in green.  }
    \label{vin_vout_alpha5}
\end{figure}

In Figure \ref{vin_vout_alpha5} we show $v_{in}-v_{out}$ curves for $\alpha=5$ and $\beta=1$. For this case, since $\alpha>4$ we have $\lambda$ real, and expect some similarity with the case of the pure $\phi^4$ model ($\beta=0$). Indeed, the structure is similar to the fractal pattern we see in the $\phi^4$ case, with three-bounce windows at the edges of the two bounce windows. However, we were not able to find three-bounce windows to the right of the two-bounce windows. These should, presumably, emerge as $\alpha$ gets larger, or as $\beta$ gets smaller. However, it is an open 
question requiring further systematic investigation
how the self-similar (on both sides) picture of
the pure $\phi^4$ model arises.

\begin{figure}[h!]
  \includegraphics[width=0.3\textwidth]{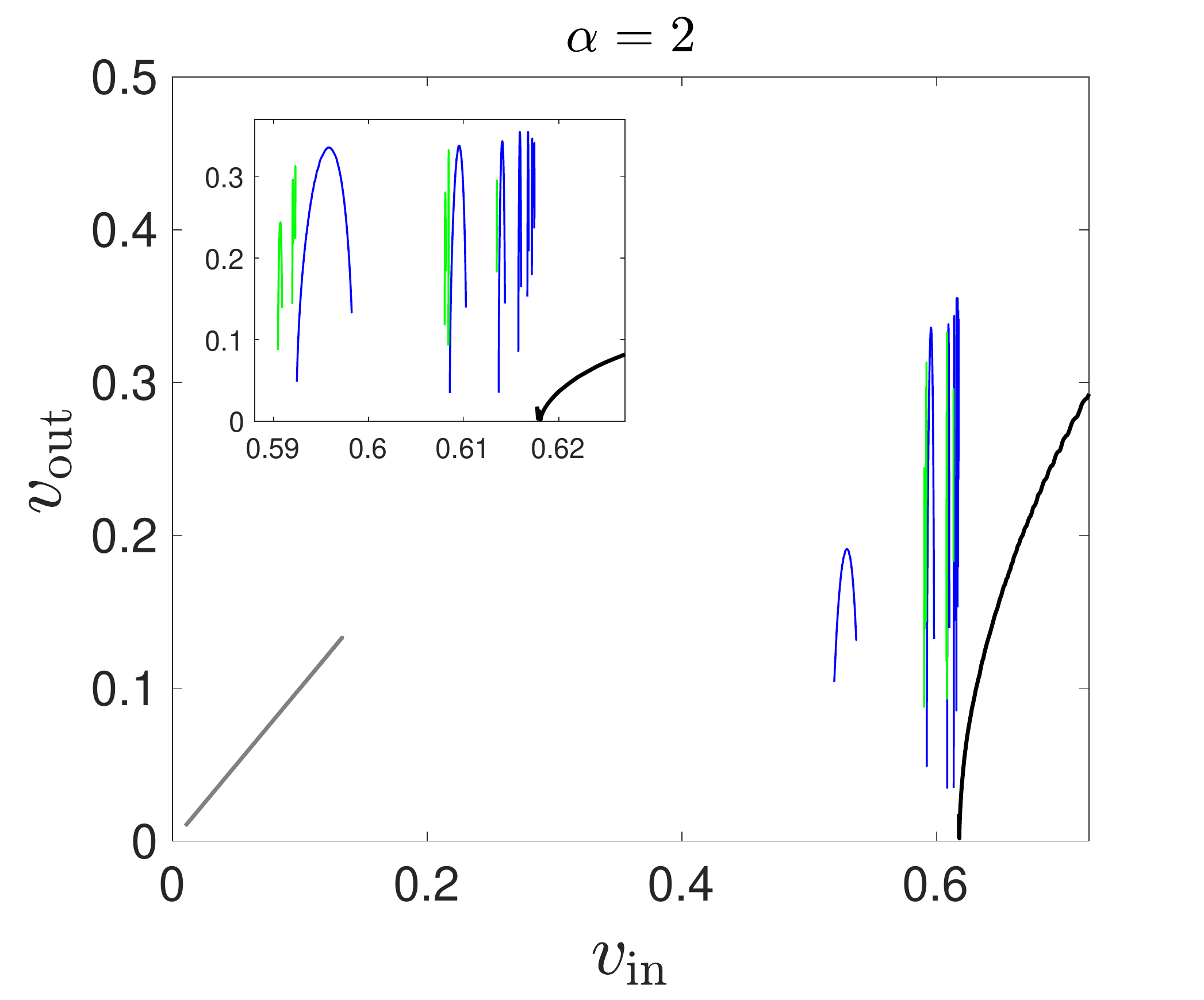}
  \includegraphics[width=0.3\textwidth]{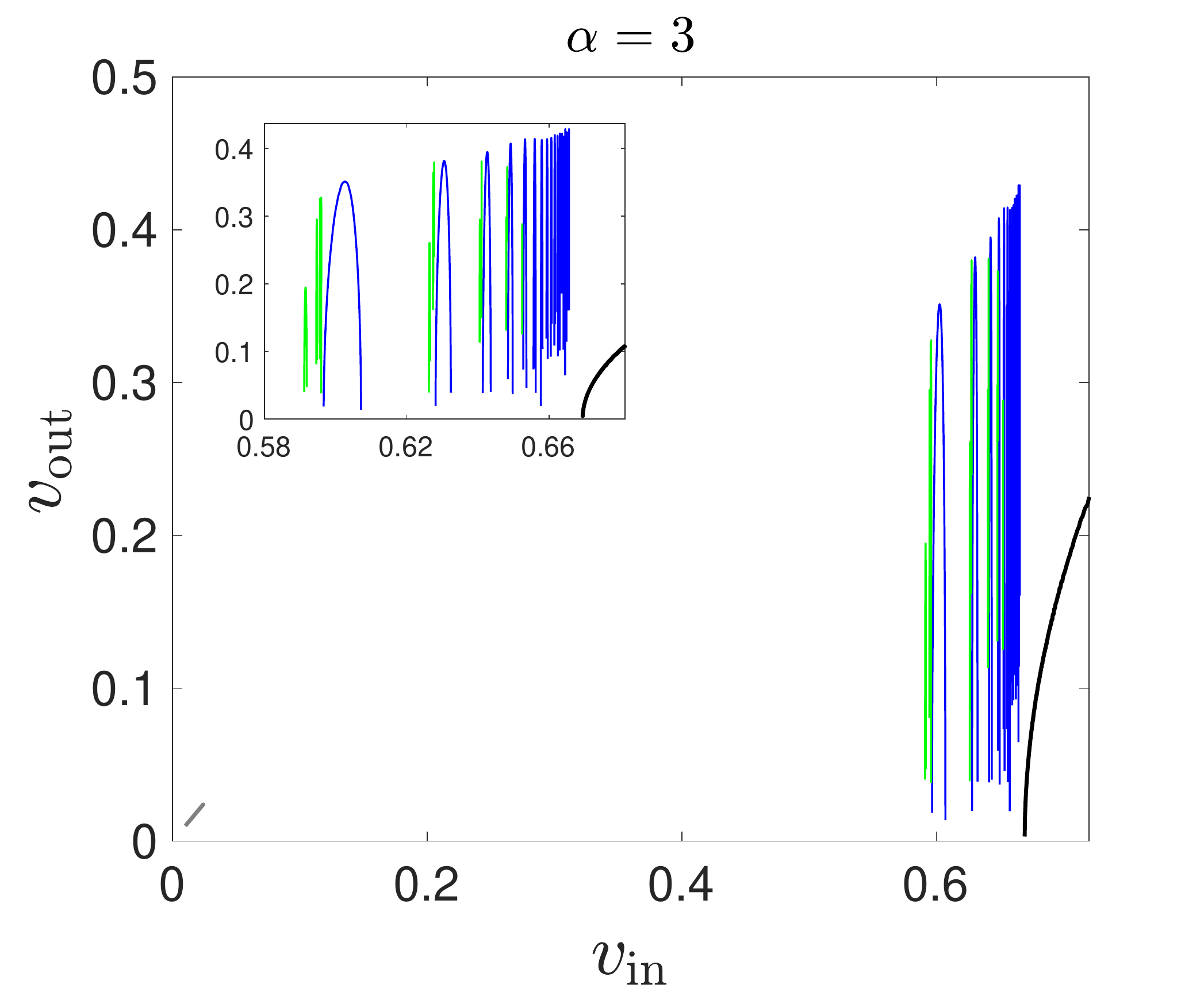}
    \includegraphics[width=0.3\textwidth]{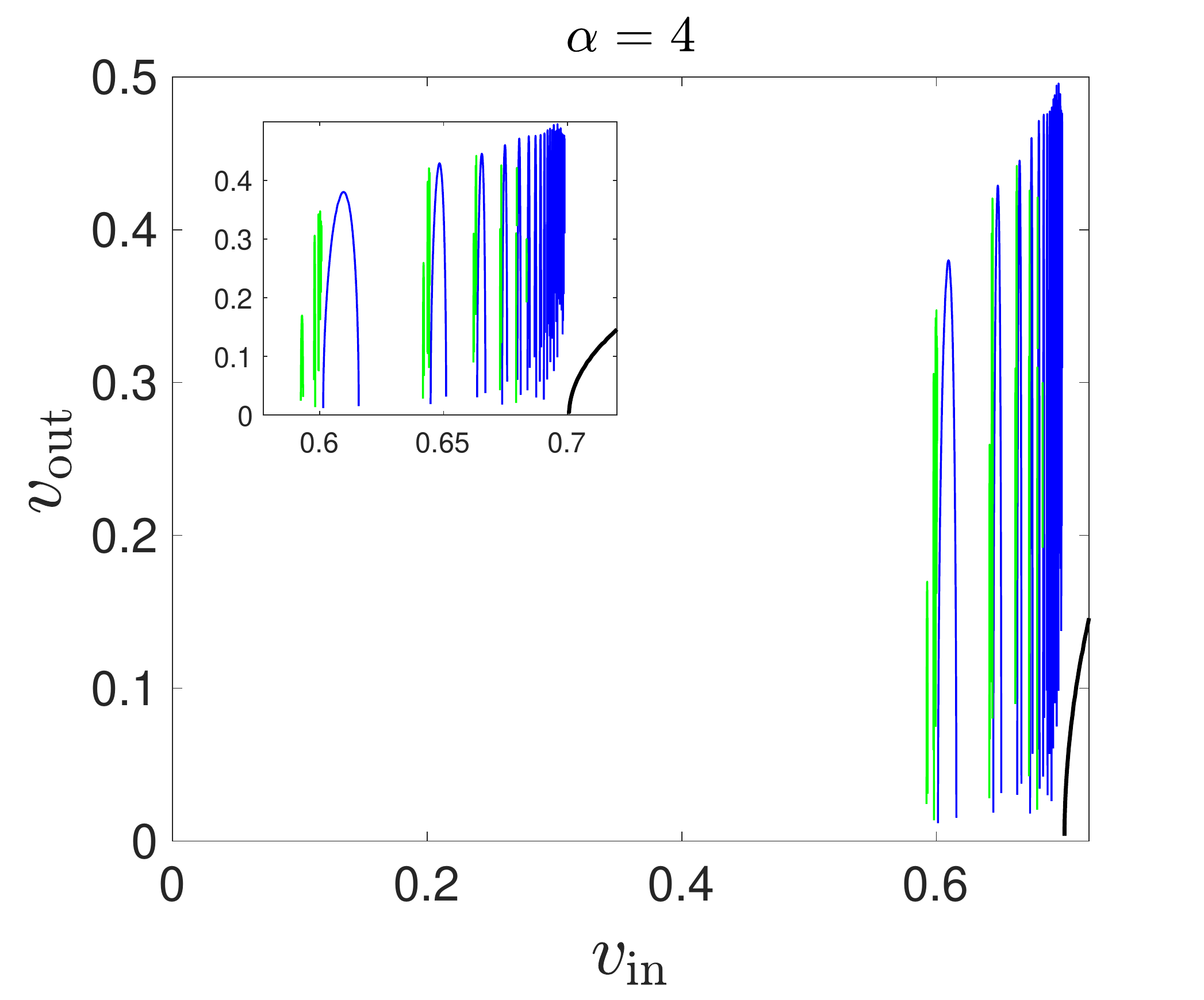}
    \caption{Transition from dominant quartic 
    progressively closer to dominant harmonic behavior, by changing $\alpha$ from $2$ (left) to $3$ (middle) and finally
    the critical case of $\alpha=4$ (right panel).}
    \label{transitionToHarmonic}
\end{figure}

We can begin to see how the system transitions from the $\alpha=0$, $\beta=1$ (pure quartic dispersion) case to the $\alpha=5$, $\beta=1$ (harmonic term dominant) case 
through some additional (less detailed) $v_{in}$-$v_{out}$ graphs; see,
e.g., Figure \ref{transitionToHarmonic}. The first two-bounce window for $\alpha=2$ (left panel, main figure, and see also Figure \ref{vin_vout} top left panel for $\alpha=1$) demonstrates a curious behavior. It appears (out of nowhere) at about $\alpha=0.711$, persisting to about $\alpha=2.8$ where it disappears.
This is why it arises in the left
panel, but not the middle one. Similarly,
notice how the transition shrinks progressively the
size of the gray line interval of ``no collision'' for $0<v_{in} <v_{1,crit}$.
It can be seen that this interval eventually disappears
for $\alpha=4$ in the right panel of the figure, 
again showcasing how the transition between the
two regimes (biharmonic vs. harmonic) emerges.

\begin{figure}[h!]
   \includegraphics[width=0.4\textwidth]{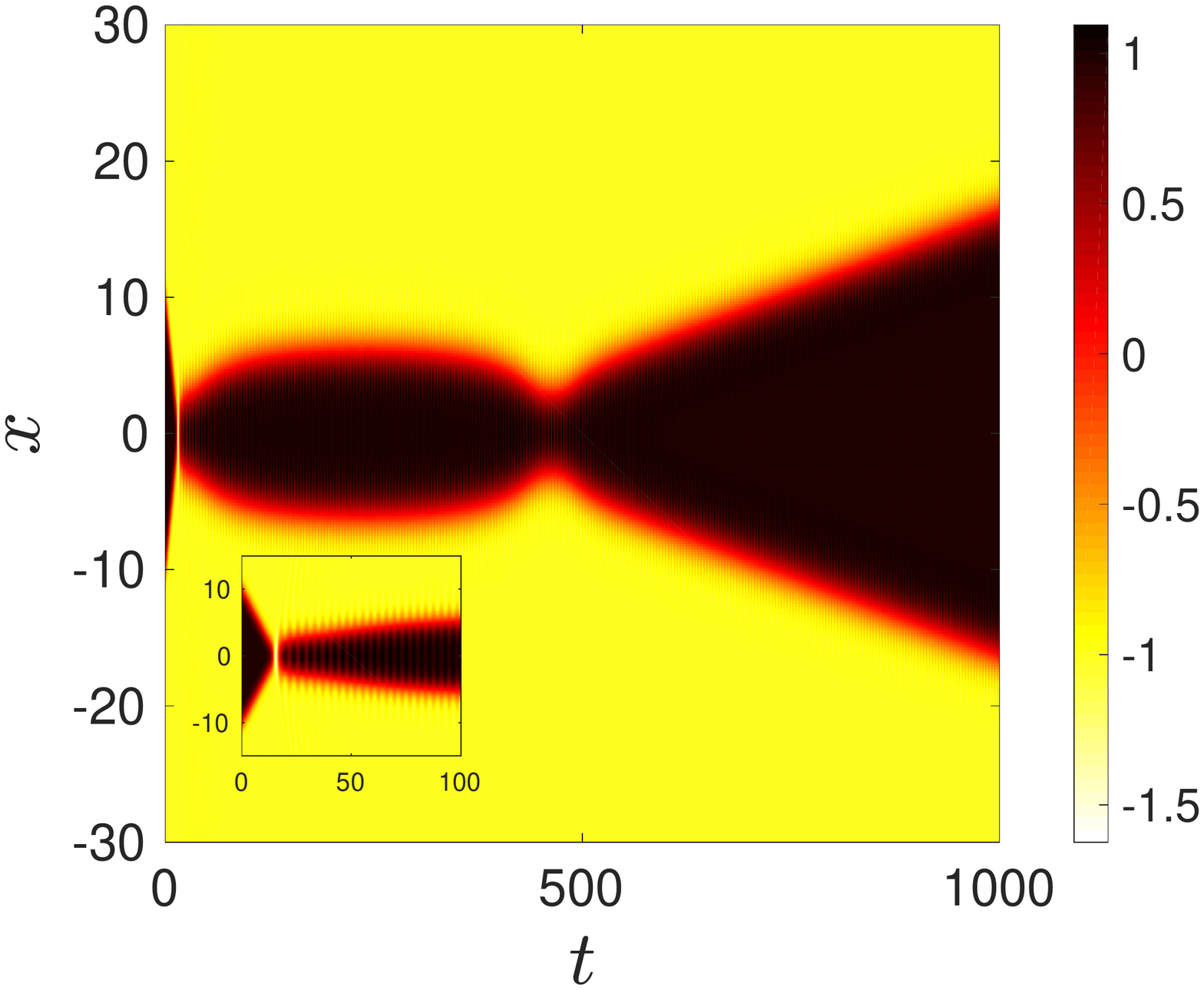}
    \includegraphics[width=0.4\textwidth]{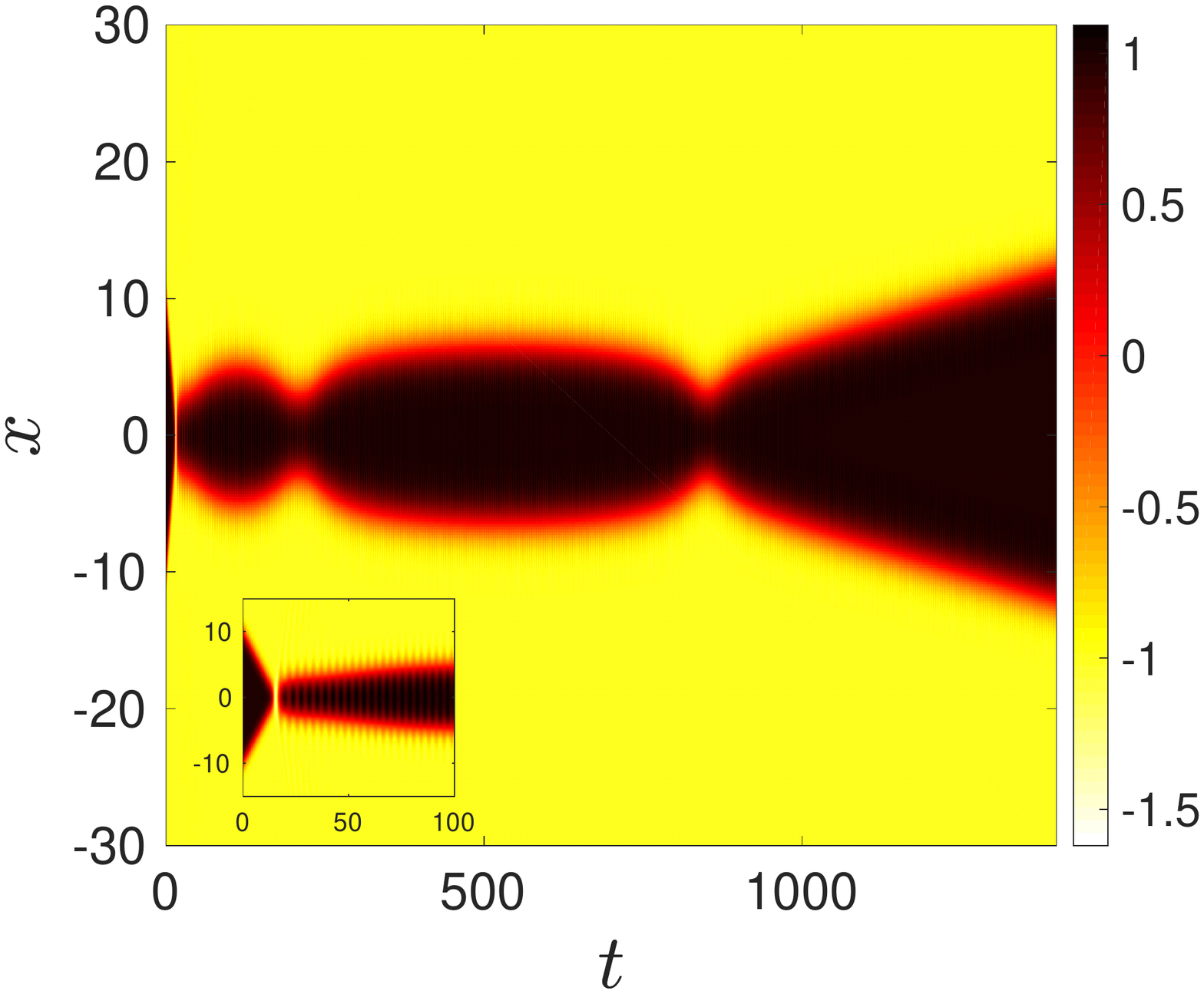}
    \caption{Contour plots when $\alpha=2.05$ and $\beta=1$ with $v_c \approx 0.6222$ for $X(0)=10$. Left panel is when $v_{\mathrm{in}}=0.622$ and right panel is when  $v_{\mathrm{in}}=0.621$.  }
    \label{newTwoBounceType}
\end{figure}

While in the discussion above, we have focused
on features that transition the phenomenology
between the two limits, it is important to realize
that the wealth of the model considered here
transcends that of solely the limit cases.
For instance,
Figure \ref{newTwoBounceType} illustrates a phenomenon not seen in either the pure biharmonic or the pure harmonic $\phi^4$ case, with $\alpha=2.05$ and $\beta=1$. What we see here is an initial interaction between kink and antikink, followed by separation of the solitons for a period of time, and then another approach of the pair. At this point one or more elastic collisions can occur, resulting in the appearance of multiple bounces. In the first panel of Figure \ref{newTwoBounceType} we see a ``pseudo'' two-bounce result, and in the second panel a pseudo three-bounce result. This can occur when the speed at which the kink and antikink approach each other for the second (or third) time is very small and therefore when we find ourselves
in the small-speed reflection window of the complex eigenvalue case.
In short, this is an unprecedented type of two-bounce since two-bounces cannot happen
in the pure biharmonic case (where there is only bion formation and single
bounce events~\cite{beam1}), but it can also not happen for pure harmonic $\phi^4$
where the small speed reflection scenario is absent. This is yet another
manifestation of the rich phenomenology of the model combining harmonic
and biharmonic dispersion. 

\section{Conclusions \& Future Challenges}

In this work we have explored a model featuring the competition of
a harmonic and biharmonic linear operator in a quadratic-quartic
$\phi^4$ model. We have argued that this model is of intrinsic
mathematical interest due to the distinct implications of the different
linear operators and also the unique features 
created by their interplay that neither of
the ``pure'' (quadratic or quartic dispersion) models
possesses. The harmonic part creates a saddle point in the
spatial dynamics and hence leads to exponentially decaying waveforms.
On the other hand, the biharmonic operator leads to complex
eigenvalues and a spiral point in the corresponding spatial dynamics.
Here, we have seen the interplay of these two possibilities creating 
an effective competition between the two tendencies. We have 
observed that this competition leads to a critical point (with an intriguing
behavior in its own right, i.e., a linearly modulated exponential)
and on the two sides of this criticality either the biharmonic
oscillatory effect or the harmonic exponential decay effect prevail
respectively. This crucially affects the interactions between the
kinks which we have also explicitly identified and corroborated
by means of detailed comparison of both the single wave tails and
of the two-coherent-structure interactions. We have also elucidated
the extent to which this collective coordinate approach can be
reliably used and illustrated its failure when the kinks get too
close to each other. Additionally, we have examined the collisions,
bounce and multi-bounce windows stemming from the kink interactions
and have shown how the critical velocities and corresponding windows
are modified as a function of the quadratic-quartic model parameters.

These findings will clearly have a significant bearing on the 
corresponding quadratic-quartic NLS model which is a natural 
possibility for future work, given the recent developments
in pure quartic solitons~\cite{pqs,pqs3} and on optical media
featuring quadratic and quartic dispersion~\cite{pqs2}. 
This natural extension will have similar existence properties
to the case considered herein, however the stability of the
dark solitons of the latter problem is an interesting open 
question. Aside from this, both at the Klein-Gordon (real 
field-theoretic) level and at the NLS one, it will be interesting
to generalize considerations to higher dimensional settings,
where it is natural to expect, e.g., vortical and more generally
topologically charged excitations. Corresponding studies are
currently in progress and will be reported in future publications.

\section*{Acknowledgments}

This material is based upon work supported by the US National Science Foundation under Grants PHY-1602994 and DMS-1809074 (P.G.K.).


\end{document}